\DocumentMetadata{}
\documentclass[sigconf,dvipsnames,anonymous=false,authorversion]{acmart}

\AtBeginDocument{%
  \providecommand\BibTeX{{%
    \normalfont B\kern-0.5em{\scshape i\kern-0.25em b}\kern-0.8em\TeX}}}

\acmConference[DaMoN '25]{21st International Workshop on Data Management on New Hardware}{June 23, 2025}{Berlin, Germany}
\usepackage{listings}
\usepackage{color}
\usepackage{xcolor}
\usepackage[plain]{algorithm2e}
\usepackage[noend]{algorithmic}
\usepackage{pifont}

\usepackage{colortbl}
\usepackage{subcaption}

\usepackage{lipsum}
\usepackage{verbatim}
\usepackage{colortbl}
\usepackage{multirow}
\usepackage{soul}

\usepackage{enumitem}

\newlist{questions}{enumerate}{2}
\setlist[questions,1]{label=\textbf{(Q\arabic*)},ref=\textbf{(Q\arabic*)}}
\setlist[questions,2]{label=(\alph*),ref=\thequestionsi(\alph*)}

\NewDocumentCommand{\codeword}{v}{%
    \texttt{\textcolor{violet}{#1}}%
}

\NewDocumentCommand{\colorcodeword}{vv}{%
    \textbf{\texttt{\textcolor{#1}{#2}}}%
}

\makeatletter
\def\NAT@def@citea{\def\@citea{\NAT@separator}}
\makeatother

\definecolor{lightgray}{gray}{0.95}

\begin{document}

\title{Bang for the Buck: Vector Search on Cloud CPUs}


% \iffalse
\author{Leonardo Kuffo}
\affiliation{%
  \institution{Centrum Wiskunde \& Informatica}
  % \institution{CWI}
  \city{Amsterdam}
  \country{The Netherlands}}
\email{lxkr@cwi.nl}

\author{Peter Boncz}
\affiliation{%
  \institution{Centrum Wiskunde \& Informatica}
  %\institution{CWI}
  \city{Amsterdam}
  \country{The Netherlands}}
\email{boncz@cwi.nl}
% \fi

%\renewcommand{\shortauthors}{Afroozeh and Kuffó and Boncz}

\begin{abstract}

Vector databases have emerged as a new type of systems that support efficient querying of high-dimensional vectors. Many of these offer their database as a service in the cloud. However, the variety of available CPUs and the lack of vector search benchmarks across CPUs make it difficult for users to choose one.  In this study, we show that CPU microarchitectures available in the cloud perform significantly differently across vector search scenarios. For instance, in an IVF index on {\small \tt float32} vectors, AMD's Zen4 gives almost 3x more queries per second (QPS) compared to Intel's Sapphire Rapids, but for HNSW indexes, the tables turn. However, when looking at the number of \textit{queries per dollar (QP\$)}, Graviton3 is the best option for most indexes and quantization settings, even over Graviton4 (Table~\ref{tab:summary}). With this work, we hope to guide users in getting the best ``bang for the buck'' when deploying vector search systems.% in the cloud.  

\end{abstract}

%\keywords{vector similarity search, approximate nearest neighbor search, data formats, vector data management, vectorized execution}

\maketitle

\section{Introduction}

The increasing amount of data in the shape of high-dimensional vectors (e.g., embeddings from large language models) has resulted in the surge of a new type of vector-centric systems known as Vector Databases (VecDBs)~\cite{vectordbs,surveysystems,llmmeetvdbs}. VecDBs like Weaviate, Milvus, Qdrant, and ChromaDB offer storage, management, and efficient querying of high-dimensional vectors~\cite{milvus, weaviate}. The demand of users for vector systems is such that existing database systems have added vector capabilities natively (MongoDB, Redis) or via extensions (e.g., pgvector in PostgreSQL, DuckDB-VSS in DuckDB). 

A core feature of VecDBs is the efficient querying of vectors through Vector Similarity Search (VSS). VSS consists of finding the vectors within a collection that are most similar to a query vector based on a distance or similarity metric. The latter is a core component of various applications, such as RAG pipelines. However, on a large scale, VSS poses challenges due to the large number of computations and storage needed to obtain exact answers to a query. To overcome this, applications gave up exactness as generally \textit{approximate} answers are ``good enough.'' The latter opened opportunities to accelerate VSS by using \textit{approximate indexes}~\cite{pqivf, hnsw, hcnng, ngt, soar}, \textit{quantization} techniques that reduce the size of the vectors~\cite{lvq, anisotropic, lowquant, rabitqext}, and optimizations to the distance evaluation~\cite{adsampling, suco, bsa, pdx}.   

%(bucket-based~\cite{pqivf, surveylsh}, tree-based~\cite{annoy, flann}, and graph-based~\cite{hnsw, nsg, hcnng, ngt}), 

% \begin{figure}[t!]
% \centering
% \includegraphics[width=0.30\columnwidth]{paper/images/dummy.png}
% \vspace*{-2mm}
% \caption{An IVF index search in the Zen4 microarchitecture gives you twice as much performance/\$ ratio. While Intel gives the best bang for your buck on HNSW indexes. Graviton4 stands as a decent multi-purpose choice.}
% \label{fig:intro}
% \vspace*{-4mm}
% \end{figure}

\begin{table}[t!]
\renewcommand{\tabcolsep}{1.5pt}
\centering
\caption{
Relative comparison of queries-per-second (QPS) and queries-per-dollar (QP\$) given by AWS cloud instances on the OpenAI/1536 dataset on various vector search scenarios. \colorbox[HTML]{A6FFA6}{++} indicates the best, \colorbox[HTML]{E9FFE9}{+} is <20\% away from the best, \colorbox[HTML]{f5f5f5}{$\cdot$} is >20\% but <2x from the best, lastly \colorbox[HTML]{fefbcb}{-} and \colorbox[HTML]{ffbfbe}{-{}-} are >2x and >3x away from the best, resp. Each (+) is 1 and each (-) is -1 score.
}
\vspace*{-4mm}
\resizebox{1\columnwidth}{!}{%
\begin{tabular}{l|l|cccccccc|ccccccccc|c|}
\cline{2-20}
                                             & \multicolumn{1}{c|}{}                                      & \multicolumn{8}{c|}{\textbf{Indexes}}                                                                                                                                                                                                             & \multicolumn{9}{c|}{\textbf{Full Scans}}                                                                                                                                                                                                                                                 &                                  \\ \cline{3-19}
                                             & \multicolumn{1}{c|}{}                                      & \multicolumn{3}{c|}{\textbf{IVF}}                                                                         & \multicolumn{5}{c|}{\textbf{HNSW}}                                                                                                             & \multicolumn{4}{c|}{\textbf{FAISS}}                                                                                                    & \multicolumn{5}{c|}{\textbf{USearch}}                                                                                                          &                                  \\
\multirow{-3}{*}{}                           & \multicolumn{1}{c|}{\multirow{-3}{*}{\textbf{Microarch.}}} & {\small f32}                        & {\small SQ}                         & \multicolumn{1}{c|}{{\small BQ}}                         & {\small f32}                        & {\small f16}                        & {\small bf}                       & {\small SQ}                         & {\small BQ}                         & {\small f32}                        & {\small SQ}                         & {\small BQ}                         & \multicolumn{1}{c|}{{\small PQ}}                         & {\small f32}                        & {\small f16}                        & {\small bf}                       & {\small SQ}                         & {\small BQ}                         & \multirow{-3}{*}{\rotatebox[origin=c]{90}{\raisebox{-0.25\normalbaselineskip}[0pt][0pt]{\bf Score}}} \\ \hline
\multicolumn{1}{|l|}{}                       & Graviton3 { \footnotesize r7g}                                              & $\cdot$                        & \cellcolor[HTML]{FFBFBE}-{}- & \multicolumn{1}{c|}{\cellcolor[HTML]{E9FFE9}+}  & \cellcolor[HTML]{E9FFE9}+  & $\cdot$                        & \cellcolor[HTML]{E9FFE9}+  & \cellcolor[HTML]{E9FFE9}+  & \cellcolor[HTML]{FEFBCB}-  & $\cdot$                        & \cellcolor[HTML]{FFBFBE}-{}- & \cellcolor[HTML]{E9FFE9}+  & \multicolumn{1}{c|}{$\cdot$}                        & $\cdot$                        & \cellcolor[HTML]{A6FFA6}++ & \cellcolor[HTML]{FFBFBE}-{}- & \cellcolor[HTML]{E9FFE9}+  & \cellcolor[HTML]{E9FFE9}+  & 2                       \\
\multicolumn{1}{|l|}{}                       & Zen3 { \footnotesize r6a}                                                   & $\cdot$                        & $\cdot$                        & \multicolumn{1}{c|}{$\cdot$}                        & \cellcolor[HTML]{E9FFE9}+  & $\cdot$                        & \cellcolor[HTML]{E9FFE9}+  & $\cdot$                        & \cellcolor[HTML]{FEFBCB}-  & $\cdot$                        & $\cdot$                        & $\cdot$                        & \multicolumn{1}{c|}{$\cdot$}                        & $\cdot$                        & $\cdot$                        & \cellcolor[HTML]{FEFBCB}-  & $\cdot$                        & $\cdot$                        & 0                                \\
\multicolumn{1}{|l|}{}                       & Graviton4 { \footnotesize r8g}                                              & $\cdot$                        & \cellcolor[HTML]{FFBFBE}-{}- & \multicolumn{1}{c|}{\cellcolor[HTML]{A6FFA6}++} & \cellcolor[HTML]{E9FFE9}+  & $\cdot$                        & \cellcolor[HTML]{E9FFE9}+  & $\cdot$                        & \cellcolor[HTML]{FEFBCB}-  & $\cdot$                        & \cellcolor[HTML]{FFBFBE}-{}- & \cellcolor[HTML]{A6FFA6}++ & \multicolumn{1}{c|}{$\cdot$}                        & $\cdot$                        & \cellcolor[HTML]{E9FFE9}+  & \cellcolor[HTML]{FEFBCB}-  & \cellcolor[HTML]{E9FFE9}+  & \cellcolor[HTML]{E9FFE9}+  & 3                                \\
\multicolumn{1}{|l|}{}                       & SPR { \footnotesize r7i}                                                    & \cellcolor[HTML]{FEFBCB}-  & $\cdot$                        & \multicolumn{1}{c|}{$\cdot$}                        & $\cdot$                        & $\cdot$                        & \cellcolor[HTML]{E9FFE9}+  & \cellcolor[HTML]{E9FFE9}+  & \cellcolor[HTML]{E9FFE9}+  & \cellcolor[HTML]{FEFBCB}-  & $\cdot$                        & $\cdot$                        & \multicolumn{1}{c|}{\cellcolor[HTML]{E9FFE9}+}  & \cellcolor[HTML]{FEFBCB}-  & $\cdot$                        & \cellcolor[HTML]{FFBFBE}-{}- & \cellcolor[HTML]{FEFBCB}-  & $\cdot$                        & -2                               \\
\multicolumn{1}{|l|}{}                       & Zen4 { \footnotesize r7a}                                                   & \cellcolor[HTML]{A6FFA6}++ & \cellcolor[HTML]{A6FFA6}++ & \multicolumn{1}{c|}{$\cdot$}                        & $\cdot$                        & \cellcolor[HTML]{FEFBCB}-  & \cellcolor[HTML]{E9FFE9}+  & $\cdot$                        & $\cdot$                        & \cellcolor[HTML]{A6FFA6}++ & \cellcolor[HTML]{A6FFA6}++ & \cellcolor[HTML]{E9FFE9}+  & \multicolumn{1}{c|}{$\cdot$}                        & \cellcolor[HTML]{A6FFA6}++ & $\cdot$                        & \cellcolor[HTML]{A6FFA6}++ & \cellcolor[HTML]{A6FFA6}++ & \cellcolor[HTML]{A6FFA6}++ & \textbf{17}                               \\
\multicolumn{1}{|l|}{\multirow{-6}{*}{\rotatebox[origin=c]{90}{\raisebox{0.40\normalbaselineskip}[0pt][0pt]{\small Queries-per-second}}}}  & SPR Z { \footnotesize r7iz}                                                 & \cellcolor[HTML]{FEFBCB}-  & $\cdot$                        & \multicolumn{1}{c|}{\cellcolor[HTML]{E9FFE9}+}  & \cellcolor[HTML]{A6FFA6}++ & \cellcolor[HTML]{A6FFA6}++ & \cellcolor[HTML]{A6FFA6}++ & \cellcolor[HTML]{A6FFA6}++ & \cellcolor[HTML]{A6FFA6}++ & \cellcolor[HTML]{FEFBCB}-  & $\cdot$                        & $\cdot$                        & \multicolumn{1}{c|}{\cellcolor[HTML]{A6FFA6}++} & \cellcolor[HTML]{FEFBCB}-  & $\cdot$                        & \cellcolor[HTML]{FEFBCB}-  & \cellcolor[HTML]{FEFBCB}-  & \cellcolor[HTML]{E9FFE9}+  & 9                                \\ \hline \hline 
\multicolumn{1}{|l|}{}                       & Graviton3 { \footnotesize r7g}                                              & \cellcolor[HTML]{A6FFA6}++ & \cellcolor[HTML]{FFBFBE}-{}- & \multicolumn{1}{c|}{\cellcolor[HTML]{A6FFA6}++} & \cellcolor[HTML]{A6FFA6}++ & \cellcolor[HTML]{A6FFA6}++ & \cellcolor[HTML]{A6FFA6}++ & \cellcolor[HTML]{A6FFA6}++ & $\cdot$                        & \cellcolor[HTML]{E9FFE9}+  & \cellcolor[HTML]{FFBFBE}-{}- & \cellcolor[HTML]{A6FFA6}++ & \multicolumn{1}{c|}{$\cdot$}                        & \cellcolor[HTML]{A6FFA6}++ & \cellcolor[HTML]{A6FFA6}++ & \cellcolor[HTML]{FEFBCB}-  & \cellcolor[HTML]{A6FFA6}++ & \cellcolor[HTML]{E9FFE9}+  & \textbf{17}                      \\
\multicolumn{1}{|l|}{}                       & Zen3 { \footnotesize r6a}                                                   & $\cdot$                        & $\cdot$                        & \multicolumn{1}{c|}{$\cdot$}                        & \cellcolor[HTML]{E9FFE9}+  & $\cdot$                        & \cellcolor[HTML]{E9FFE9}+  & $\cdot$                        & \cellcolor[HTML]{FEFBCB}-  & \cellcolor[HTML]{E9FFE9}+  & \cellcolor[HTML]{A6FFA6}++ & $\cdot$                        & \multicolumn{1}{c|}{$\cdot$}                        & \cellcolor[HTML]{E9FFE9}+  & $\cdot$                        & $\cdot$                        & $\cdot$                        & $\cdot$                        & 5                                \\
\multicolumn{1}{|l|}{}                       & Graviton4 { \footnotesize r8g}                                              & $\cdot$                        & \cellcolor[HTML]{FFBFBE}-{}- & \multicolumn{1}{c|}{\cellcolor[HTML]{E9FFE9}+}  & \cellcolor[HTML]{E9FFE9}+  & \cellcolor[HTML]{E9FFE9}+  & \cellcolor[HTML]{A6FFA6}++ & \cellcolor[HTML]{E9FFE9}+  & $\cdot$                        & \cellcolor[HTML]{E9FFE9}+  & \cellcolor[HTML]{FEFBCB}-  & \cellcolor[HTML]{E9FFE9}+  & \multicolumn{1}{c|}{$\cdot$}                        & \cellcolor[HTML]{E9FFE9}+  & \cellcolor[HTML]{E9FFE9}+  & $\cdot$                        & \cellcolor[HTML]{E9FFE9}+  & \cellcolor[HTML]{A6FFA6}++ & 10                               \\
\multicolumn{1}{|l|}{}                       & SPR { \footnotesize r7i}                                                    & \cellcolor[HTML]{FEFBCB}-  & $\cdot$                        & \multicolumn{1}{c|}{$\cdot$}                        & $\cdot$                        & \cellcolor[HTML]{E9FFE9}+  & \cellcolor[HTML]{E9FFE9}+  & $\cdot$                        & \cellcolor[HTML]{A6FFA6}++ & \cellcolor[HTML]{FEFBCB}-  & $\cdot$                        & \cellcolor[HTML]{FEFBCB}-  & \multicolumn{1}{c|}{\cellcolor[HTML]{A6FFA6}++} & \cellcolor[HTML]{FEFBCB}-  & \cellcolor[HTML]{FEFBCB}-  & \cellcolor[HTML]{FEFBCB}-  & \cellcolor[HTML]{FEFBCB}-  & $\cdot$                        & -1                               \\
\multicolumn{1}{|l|}{}                       & Zen4 { \footnotesize r7a}                                                   & \cellcolor[HTML]{A6FFA6}++ & \cellcolor[HTML]{A6FFA6}++ & \multicolumn{1}{c|}{$\cdot$}                        & $\cdot$                        & \cellcolor[HTML]{FEFBCB}-  & $\cdot$                        & $\cdot$                        & $\cdot$                        & \cellcolor[HTML]{A6FFA6}++ & \cellcolor[HTML]{A6FFA6}++ & $\cdot$                        & \multicolumn{1}{c|}{$\cdot$}                        & \cellcolor[HTML]{A6FFA6}++ & \cellcolor[HTML]{FEFBCB}-  & \cellcolor[HTML]{A6FFA6}++ & $\cdot$                        & $\cdot$                        & 10                               \\
\multicolumn{1}{|l|}{\multirow{-6}{*}{\rotatebox[origin=c]{90}{\raisebox{0.40\normalbaselineskip}[0pt][0pt]{\small Queries-per-dollar}}}} & SPR Z { \footnotesize r7iz}                                                 & \cellcolor[HTML]{FEFBCB}-  & \cellcolor[HTML]{FEFBCB}-  & \multicolumn{1}{c|}{$\cdot$}                        & $\cdot$                        & \cellcolor[HTML]{E9FFE9}+  & $\cdot$                        & $\cdot$                        & \cellcolor[HTML]{E9FFE9}+  & \cellcolor[HTML]{FFBFBE}-{}- & $\cdot$                        & \cellcolor[HTML]{FEFBCB}-  & \multicolumn{1}{c|}{$\cdot$}                        & \cellcolor[HTML]{FEFBCB}-  & \cellcolor[HTML]{FEFBCB}-  & \cellcolor[HTML]{FFBFBE}-{}- & \cellcolor[HTML]{FFBFBE}-{}- & $\cdot$                        & -9                               \\ \hline
\end{tabular}
}
\label{tab:summary}
\vspace*{-4mm}
\end{table}

%\multirow{-3}{*}{\rotatebox[origin=c]{90}{\raisebox{-0.25\normalbaselineskip}[0pt][0pt]{\bf Score}}}

% \footnotesize

% -> $\cdot$
% -{}- -> -{}-

A typical business model of VecDBs is to offer their software as a service (SaaS), either in a cloud environment owned by them or by the users (``Bring Your Own Cloud''). This situation prompts the question: \textit{Which cloud instance is the best for vector search?} While most VecDBs can be deployed on both major architectures (x86\_64 and ARM), there is a lack of benchmarks comparing the performance of vector search across different architectures, let alone microarchitectures (e.g., Intel Sapphire Rapids, AMD Zens, AWS Gravitons). %Usually, VecDB vendors recommend having sufficient DRAM to fit the vectors. 
Qdrant, in particular, strongly advises using the latest-generation Intel processors~\cite{intelqdrant}. Milvus, Chroma, and Vexless have presented benchmarks of their systems on Intel CPUs~\cite{milvusavx,chromaperf,vexless}, while USearch~\cite{usearch} presents benchmarks on AWS Graviton3 (ARM). 

However, far from being as trivial as choosing the CPUs with the largest caches, highest clock frequency, or specialized SIMD instructions, we uncover that the optimal choice depends on the search algorithm and quantization level used. For instance, in partition-based indexes, like IVF~\cite{pqivf}, AMD's Zen4 gives almost 3x more queries per second (QPS) than Intel's Sapphire Rapids, but the tables turn on graph indexes, like HNSW~\cite{hnsw}, in which Intel Sapphire Rapids delivers more QPS. However, when looking at the number of queries per dollar (QP\$), AWS Graviton3 gives the best bang for the buck, even over its successor, Graviton4. 

% This study aims to show which cloud CPUs give the best ``bang for the buck'' by experimentally evaluating their QPS and QP\$ on different vector search scenarios. More importantly, we uncover that the search algorithm's data-access patterns plays an important role in the performance differences across microarchitectures. The latter makes the CPU cache performance (bandwidth and latency) important for vector search, as it is memory-bound in many scenarios~\cite{pdx, sancaaccess, graphmembounded}.

% To show which cloud CPUs gives the best ``bang for the buck", 
This study aims to show which cloud CPUs give the best ``bang for the buck'' by experimentally evaluating their QPS and QP\$ on different vector search scenarios. More importantly, we uncover that the performance across microarchitectures depends not solely on their SIMD capabilities but also on the search algorithm's data-access patterns. The latter makes the CPU cache performance (bandwidth and latency) important for vector search, as it is memory-bound in many scenarios~\cite{pdx, sancaaccess, graphmembounded, scannwhitepaper}. % Another insight from our study is that while some microarchitectures provide SIMD for particular data types (e.g., {\small \tt POPCOUNT} in Intel Sapphire Rapids for 1-bit vectors), they do not always deliver the best QP\$. 

%Finally, we also identify certain microarchitectures that deliver significantly lower performance under certain scenarios (e.g. Graviton4 with 6-bit vectors delivers 5x less QP\$ than in Zen4).

%We found that by choosing the right microarchitecture, one can get up to . 

% In this work, we experimentally evaluate the QP\$ of 5 CPU microarchitectures (AMD Zen4 / Zen3, Intel Sapphire Rapids, and AWS Graviton4 / Graviton3) on searches within vector indexes (IVF and HNSW) and full scans, quantized at different levels. Finally, we give an in-depth analysis on the \textit{why} vector search algorithms perform differently across microarchitectures. 

% In this work, we make the following contributions: \textbf{(i)} A guide to help users determine which microarchitecture to choose based on their use case. \textbf{(ii)} An experimental evaluation of the QP\$ of 4 CPU microarchitectures (AMD Zen4/Zen3, Intel Sapphire Rapids, and Graviton4) on searches within IVF and HNSW indexes quantized at different levels, using their current on-demand price in AWS us-east-1. \textbf{(iii)} An in-depth analysis on the \textit{why} vector search algorithms perform differently across microarchitectures. 

\section{Preliminaries}\label{sec:preliminaries}

\subsection{Approximate Vector Similarity Search}\label{sec:theknnproblem}
Given a collection $V$ of $n$ multi-dimensional objects $\{v_0, v_1, \cdots, v_{n-1}\}$, defined on a $D$-dimensional space, and a $D$-dimensional query $q$, VSS tries to find the subset $R \subset V$, containing the $k$ most \textit{similar} vectors to $q$. The notion of similarity between two vectors $(v, q)$ is measured using a function $\delta(v, q)$. Usually, $\delta$ is a distance or similarity function defined in an Euclidean space. The Squared Euclidean Distance (L2) is one of the most commonly used distance metrics, and it is defined as $\delta(v, q) = \sum_{i=0}^{D} (v_i - q_i)^2$. To obtain $R$, $\delta$ must be computed for every $v\in V$, leading to a large number of calculations. However, in most vector-based applications, \textit{approximate} answers are acceptable. This allowed VSS to scale by returning only an approximate result set $\hat{R}$, whose quality is measured by the percentage of intersection between the vectors in $R$ and $\hat{R}$ when answering the same query (\textit{recall} metric). This tradeoff between accuracy and speed resulted in the development of approximate indexes and quantization techniques for vectors.

\begin{figure}[t!]
\centering
\includegraphics[width=1.0\columnwidth]{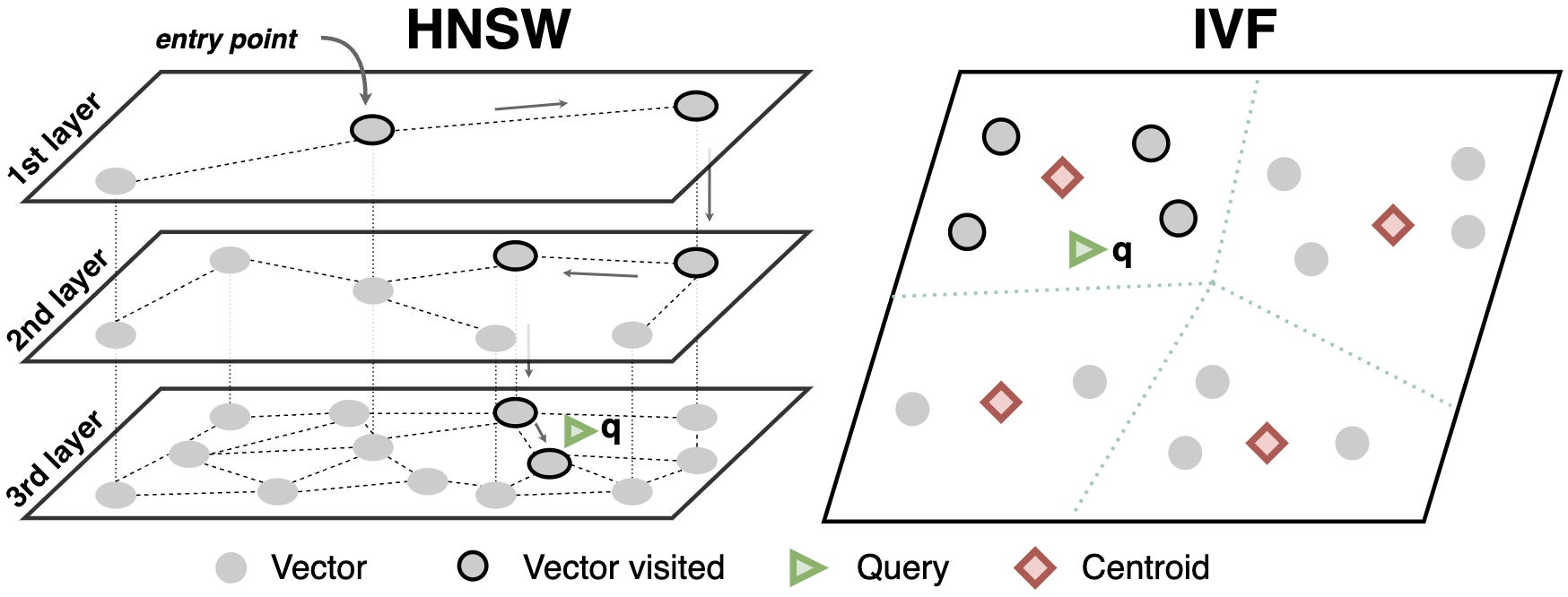}
%\vspace*{-4mm}
\caption{Example of an HNSW and IVF index search.}
%\vspace*{-6mm}
\label{fig:ivf-hnsw}
\end{figure}

\subsection{Approximate Indexes}
Approximate indexes aim to build data structures that guide the search to the most suitable place of the D-dimensional space in which the query \textit{may} find its nearest neighbours. Indexes can be categorized into three types: graph-based~\cite{hnsw, hcnng, surveygraph, patel2024acorn}, partition-based~\cite{annoy, pqivf, surveylsh}, and hybrids~\cite{spann, ngt}. Their common goal is only to evaluate the distance/similarity function between $q$ and a smaller set of vectors $V' \subset V$ while maintaining high recalls. The indexes that have seen the most adoption in vector systems are HNSW (Hierarchical Navigable Small Worlds)~\cite{hnsw} and IVF (Inverted Files)~\cite{pqivf}. %, with the possibility to quantize the vectors from 32-bit values to 16, 8, 4, 6, or even 1-bit to reduce storage requirements.

\vspace*{3mm}\noindent{\bf HNSW} has seen great success in achieving desirable recall in most datasets~\cite{annbench, candy}. HNSW organizes objects into a graph where nodes represent the vectors, and edges reflect their similarity. The property of navigability and "small world"~\cite{smallworld} is forced on the graph so that a greedy search can reach the answers to a query in logarithmic time~\cite{surveygraph}. Borrowing ideas from the skiplist data structure, HNSW organizes the nodes into different layers (see left of Figure ~\ref{fig:ivf-hnsw}). The top layer (starting point) contains ``distant" nodes, and the bottom layer has all the nodes. The upper layer allows a search to quickly traverse the graph diameter with a few steps, and the lower layer allows it to traverse on a local hub of nodes. 

On the other hand, {\bf IVF} is a partition-based index that clusters the vector collection into buckets. At search time, the distance metric is first evaluated with the centroids of each bucket, and the vectors inside the nearest centroids buckets are chosen for evaluation (see right of Figure ~\ref{fig:ivf-hnsw}). The number of centroids is set in the order of $\sqrt{n}$~\cite{faisspaper, milvus}. More buckets can be probed to trade off speed for more recall~\cite{faisspaper, milvus}. IVF works modestly well in most datasets~\cite{annbench, candy} while scaling better than graph indexes, which have higher memory requirements and longer construction times~\cite{faisspaper, scannwhitepaper}. A commonly used \textit{hybrid index} consists of building an HNSW index on the IVF centroids to quickly find the most promising buckets~\cite{spann}.

\begin{figure}[t!]
\centering
\includegraphics[width=1\columnwidth]{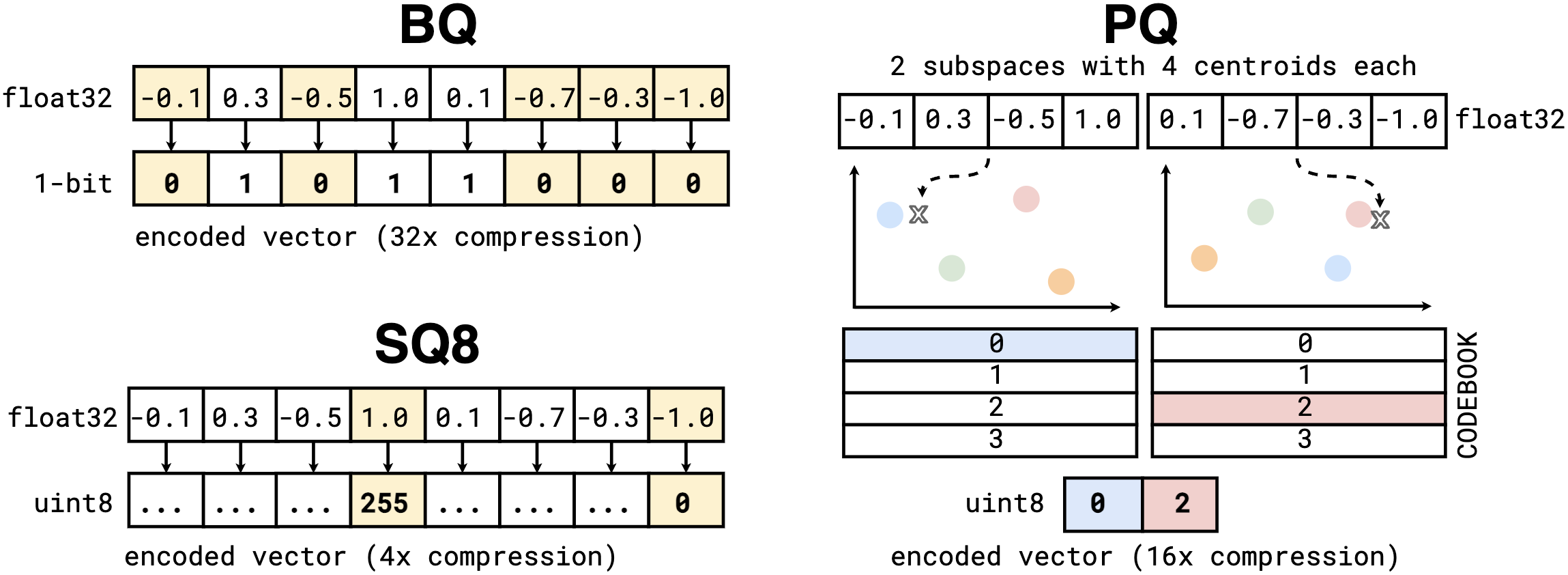}
% \vspace*{-4mm}
\caption{Overview of different quantization techniques.}
%\vspace*{-4mm}
\label{fig:quantization}
\end{figure}

\subsection{Quantization Techniques}
Quantization techniques aim to reduce the size of every vector. Quantization can be used on top of indexes to alleviate storage requirements. However, the more aggressive quantization, the more the recall is reduced as precision is lost. Quantization can increase query throughput, as less data must be fetched~\cite{sancaaccess}, and \textit{SIMD} can operate on more values at-a-time with a single CPU instruction. % However, depending on the available SIMD intrinsics, and the implementation of the distance kernels, the search performance may also \textit{decrease}. 

Among the most popular quantization techniques, we have binary quantization (BQ), scalar quantization (SQ), and product quantization (PQ)~\cite{pqivf}. Examples of these techniques are shown in Figure ~\ref{fig:quantization}. BQ maps every {\tt \small float32} value in the vector to 0 or 1 (32x compression). In Figure ~\ref{fig:quantization}, positive values are mapped to 1 and negative values to 0. In SQ, each {\tt \small float32} value is mapped to an integer code of usually 8, 6, or 4-bit, depending on the target compression ratio. In Figure ~\ref{fig:quantization}, each value is mapped to 8 bits by normalizing and scaling them to the [0-255] range. The latter is one of the simplest ways to perform SQ. However, a wider family of SQ algorithms exists~\cite{rabitqext, lvq}. Finally, \textit{downcasting} to {\small\tt float16} or {\small\tt bfloat} is a safer alternative with little information loss but only achieving a 2x compression ratio.

In PQ, the D-dimensional space is split into $M$, $\frac{D}{M}$ dimensional subspaces, and a codebook is trained for each subspace using a clustering algorithm. In Figure ~\ref{fig:quantization}, we split vectors into two subspaces of $\frac{8}{2}=4$ dimensions and create 4 clusters for each subspace. To encode a vector, PQ splits it into $M$ subvectors. For each subvector, PQ finds the nearest centroid on that subspace codebook, and the vector is encoded with the centroid codes. Contrary to BQ and SQ, PQ can provide variable compression ratios depending on the chosen number of subspaces and centroids per subspace. However, contrary to BQ and SQ, in PQ, the distance calculations cannot happen in the quantized domain, as each code must be first decoded. The latter affects the distance calculation latency. A wider family of PQ algorithms exists, mostly improving how the codebook is constructed to tradeoff search speed, size, and recall~\cite{anisotropic,faisspaper}. 

\vfill

\begin{table*}[t!]
\renewcommand{\tabcolsep}{3.0pt}
\centering
\caption{AWS cloud instances used (May 2025, us-east-1)}
\vspace*{-4mm}
\resizebox{1\linewidth}{!}{%
\begin{tabular}{llccccccccccccc}
\hline
\multirow{2}{*}{\textbf{Microarchitecture}} & \multirow{2}{*}{\textbf{CPU Model}} & \multirow{2}{*}{\textbf{\begin{tabular}[c]{@{}c@{}}Instance\\ Name\end{tabular}}} & \multirow{2}{*}{\textbf{\begin{tabular}[c]{@{}c@{}}Price/h\\ $\$$USD $\downarrow$ \end{tabular}}} & \multirow{2}{*}{\textbf{\begin{tabular}[c]{@{}c@{}}Freq.\\ GHz\end{tabular}}} & \multirow{2}{*}{\textbf{\begin{tabular}[c]{@{}c@{}}L1\\ KiB\end{tabular}}} & \multirow{2}{*}{\textbf{\begin{tabular}[c]{@{}c@{}}L2\\ MiB\end{tabular}}} & \multirow{2}{*}{\textbf{\begin{tabular}[c]{@{}c@{}}L3\\ MiB\end{tabular}}} & \multirow{2}{*}{\textbf{\begin{tabular}[c]{@{}c@{}}SIMD \\ ISA\end{tabular}}} & \multirow{2}{*}{\textbf{\begin{tabular}[c]{@{}c@{}}SIMD \\ Width\end{tabular}}} & \multicolumn{5}{c}{\textbf{SIMD Capabilities}} \\
                                            &                                     &                                                                                   &                                                                                   &                                                                               &                                                                            &                                                                            &                                                                            &                                                                               &                                                                                 & {\tt \small f32}     & {\tt \small f16}     & {\tt \small bf}      & {\tt \small i32,16,8}   & {\tt \small popcnt}  \\ \hline
AWS Graviton 3                              & Neoverse V1                         & r7g.2x                                                                            & 0.4284                                                                            & 2.6                                                                           & 64                                                                         & 1                                                                          & 32                                                                         & NEON / SVE                                                                    & 128 / 256                                                                       & \ding{51}   & \ding{51}   & \ding{51}   & \ding{51}  & \ding{51}   \\
AMD Zen 3                                   & EPYC 7R13                           & r6a.2x                                                                            & 0.4536                                                                            & 3.6                                                                           & 32                                                                         & 0.5                                                                        & 16                                                                         & AVX2                                                                          & 256                                                                             & \ding{51}   & \ding{55}   & \ding{55}   & \ding{51}  & \ding{55}   \\
AWS Graviton 4                              & Neoverse V2                         & r8g.2x                                                                            & 0.4713                                                                            & 2.8                                                                           & 64                                                                         & 2                                                                          & 36                                                                         & NEON / SVE2                                                                   & 128 / 128                                                                       & \ding{51}   & \ding{51}   & \ding{51}   & \ding{51}  & \ding{51}   \\
Intel Sapphire Rapids                       & Platinum 8488C                      & r7i.2x                                                                            & 0.5292                                                                            & 3.2                                                                           & 48                                                                         & 2                                                                          & 105                                                                        & AVX512                                                                        & 512                                                                             & \ding{51}   & \ding{51}   & \ding{51}   & \ding{51}  & \ding{51}   \\
AMD Zen 4                                   & EPYC 9R14                           & r7a.2x                                                                            & 0.6086                                                                            & 3.7                                                                           & 32                                                                         & 1                                                                          & 32                                                                         & AVX512                                                                        & 512                                                                             & \ding{51}   & \ding{55}   & \ding{51}   & \ding{51}  & \ding{51}   \\
Intel Sapphire Rapids Z                     & Gold 6455B                          & r7iz.2x                                                                           & 0.7440                                                                            & 3.9                                                                           & 48                                                                         & 2                                                                          & 60                                                                         & AVX512                                                                        & 512                                                                             & \ding{51}   & \ding{51}   & \ding{51}   & \ding{51}  & \ding{51}   \\ \hline
\end{tabular}
}
\label{tab:machine_details}
%\vspace*{-2mm}
\end{table*}

\subsection{Single Instruction Multiple Data (SIMD)}
Distance calculations can be optimized in CPUs using SIMD intrinsics that process multiple values with a single CPU instruction. 

\vspace*{3mm}\noindent{\bf SIMD in x86\_64 and ARM.} In x86\_64 architectures, SIMD instructions are called Advanced Vector Extensions (AVX). The number of values AVX can process at a time depends on the SIMD register width supported by the CPU. Initially, registers of 256-bit were introduced (AVX and AVX2), further expanded to 512-bit (AVX512), which can process 16 {\small\tt float32} values with one instruction. AVX512 had a rough start, reportedly down-clocking the CPU in earlier Intel microarchitectures (Sky Lake and earlier)~\cite{lemireavx512bad}. However, this is no longer an issue in modern CPUs (Zen4, Sapphire Rapids), and AVX512 is widely used to accelerate vector search. On the other hand, ARM architectures also provide SIMD instructions through NEON and SVE. NEON was introduced first, supporting SIMD over 128-bit registers. SVE was introduced later as an improvement over NEON that, unlike traditional SIMD architectures, supports variable-size SIMD registers on its intrinsics through VLA (Variable Length Agnostic) programming. The latter alleviates technical debt as distance kernels no longer need hardware-dependent loop lengths. Graviton4 has 128-bit SVE registers, and Graviton3 has 256-bit SVE registers, with both having 128-bit NEON registers. 

% Figure \ref{fig:simd} shows an example of an Inner Product calculation between two {\small\tt float32} vectors using AVX512. 

\vspace*{3mm}\noindent{\bf SIMD capabilities.} SIMD instructions usually support floating-point arithmetic (double and single precision), integer arithmetic (64, 32, 16, and 8-bit), data movement, conversions, and comparisons. In addition to this, CPU vendors have added extended SIMD capabilities in different microarchitectures. For instance, Intel's Sapphire Rapids (SPR) supports arithmetic of {\small\tt float16}, and both AMD's Zen4 and Intel's SPR support arithmetic of {\small\tt bfloat} and a {\small\tt POPCOUNT} instruction useful for 1-bit vectors. In modern ARM CPUs, SVE and NEON support arithmetic for {\small\tt bfloat}, half-precision {\small\tt float16}, and 1-bit vectors ({\small \tt POPCOUNT}). Graviton4 supports SVE2, an extension to SVE that introduces additional intrinsics such as {\small \tt MATCH} to compute the intersection between two vectors. 

Table \ref{tab:machine_details} shows some of the SIMD capabilities relevant to distance calculations present in modern CPUs available in the cloud.
It is important to note that, in some scenarios, SIMD instructions are not directly usable in distance kernels. One example is trying to do a dot product between two 8-bit vectors in AVX2/AVX512, as the only available instruction to do so (i.e., {\small \tt VPDPBUSD}) expects one vector to be {\small \tt signed} and the other to be {\small \tt unsigned}. Challenges also arise with sub-byte kernels. For instance, vectors quantized to 6 bits must first be \textit{aligned} into the SIMD registers, usually with {\small \tt SHIFT+OR} instructions, which impact the performance of the distance kernels.

%\vspace*{3mm}
\noindent{\bf SIMD in vector libraries.} Some vector libraries, like USearch~\cite{usearch}, leverage SIMD capabilities to provide distance kernels for quantized types in most major CPU microarchitectures. On the other hand, other libraries, like FAISS, avoid specialized kernels by decoding vectors back to the {\small \tt float32} domain. The latter is called asymmetric distance calculations. This reduces technical debt and avoids complex code bases. However, the performance of such kernels is not on par with symmetric kernels that operate in the quantized domain~\cite{simsimd}. Both USearch and FAISS prefer to use the latest SIMD ISA available in the CPU (e.g., SVE over NEON if both are available).

\vspace*{3mm}\noindent{\bf x86\_64 vs ARM SIMD.} A key difference between the SIMD of microarchitectures lies in their register width. However, a larger register width does not guarantee better raw performance. For instance, CPUs with NEON SIMD do not fall behind AVX512 on database workloads~\cite{fastlanes} despite having a 4x smaller register width. This is because the latency and execution throughput of the instructions used also impact performance.

% \begin{figure}[t]
% \includegraphics[width=0.5\linewidth]{paper/images/simd_vs_pdx.png}
% \centering
% \vspace*{-5.5mm}
% \caption{An Inner Product calculation between two vectors using AVX512 (show here registers figure and code).}
% \label{fig:simd}
% \vspace*{-4mm}
% \end{figure}

\begin{table}[t!]
\renewcommand{\tabcolsep}{3.0pt}
\centering
\caption{Vector datasets}
\vspace*{-4mm}
\label{tab:datasets}
\resizebox{0.98\columnwidth}{!}{%
\begin{tabular}{llcccc}
\hline
\multicolumn{1}{c}{\textbf{Dataset}} & \multicolumn{1}{l}{\textbf{Semantics}} & \textbf{Size}       & \textbf{N. Queries}       &  \textbf{Dim.$\uparrow$} & \multicolumn{1}{c}{\textbf{Distribution}} \\ \hline 
OpenAI                         & Text Embeddings                 & 999,000    & 1,000    & 1536        & \raisebox{-.4\height}{\includegraphics[width=0.06\textwidth]{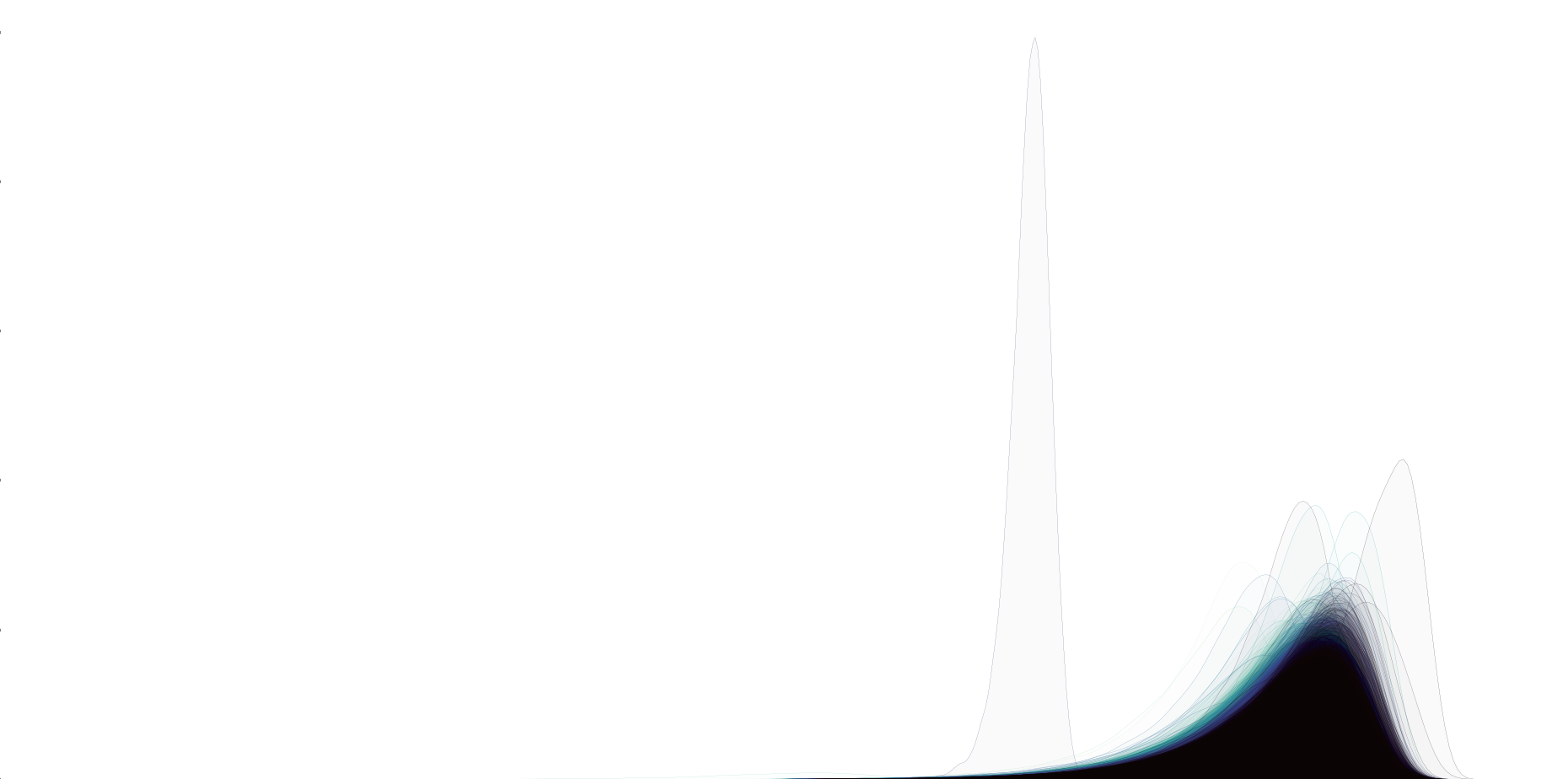}}                                  \\ 
arXiv                         & Text Embeddings                 & 2,253,000 & 1,000    & 768        & \raisebox{-.4\height}{\includegraphics[width=0.06\textwidth]{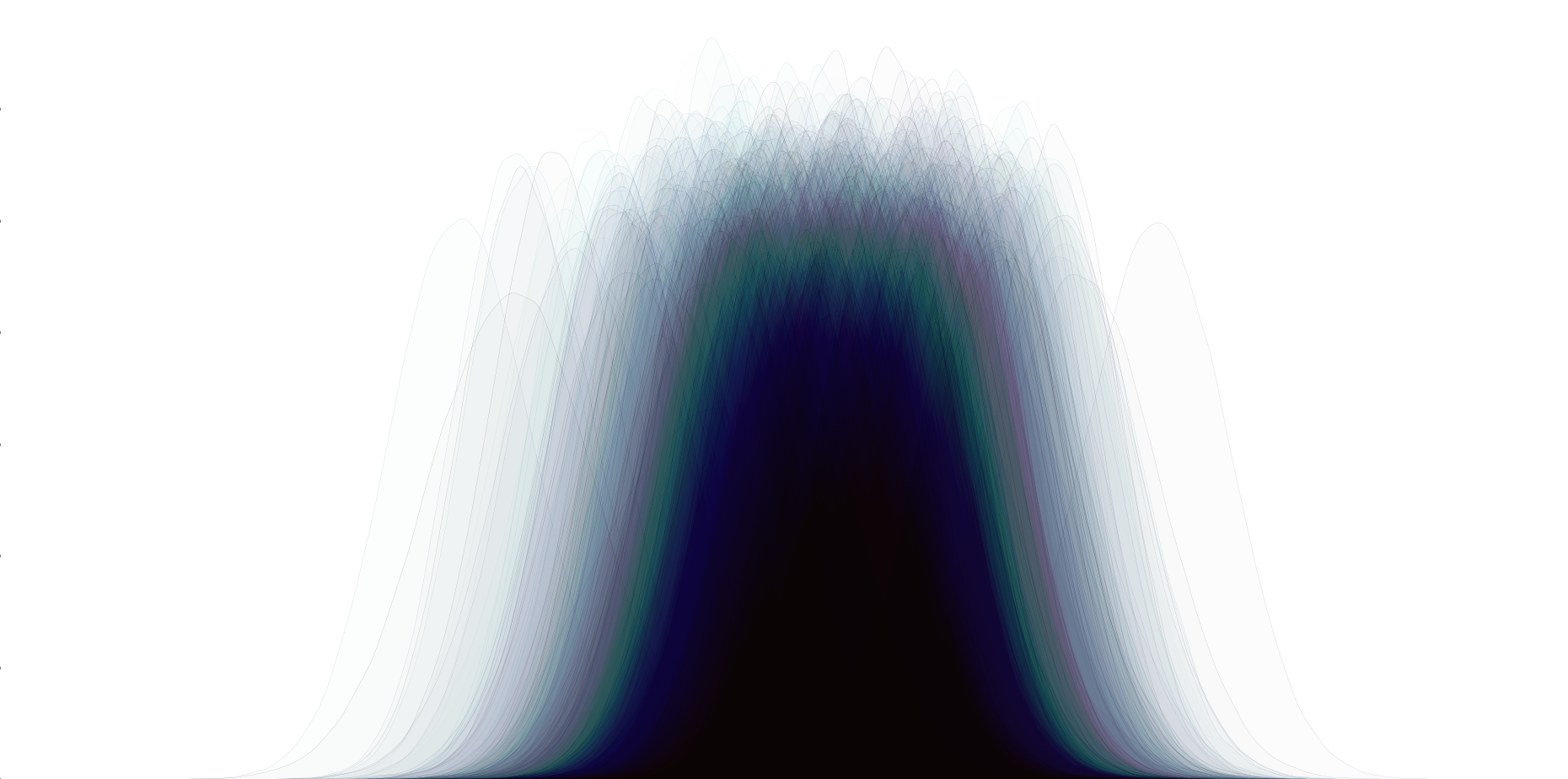}}                                  \\ 
SIFT                          & Image Features                 & 1,000,000  & 10,000  & 128        &  \raisebox{-.4\height}{\includegraphics[width=0.06\textwidth]{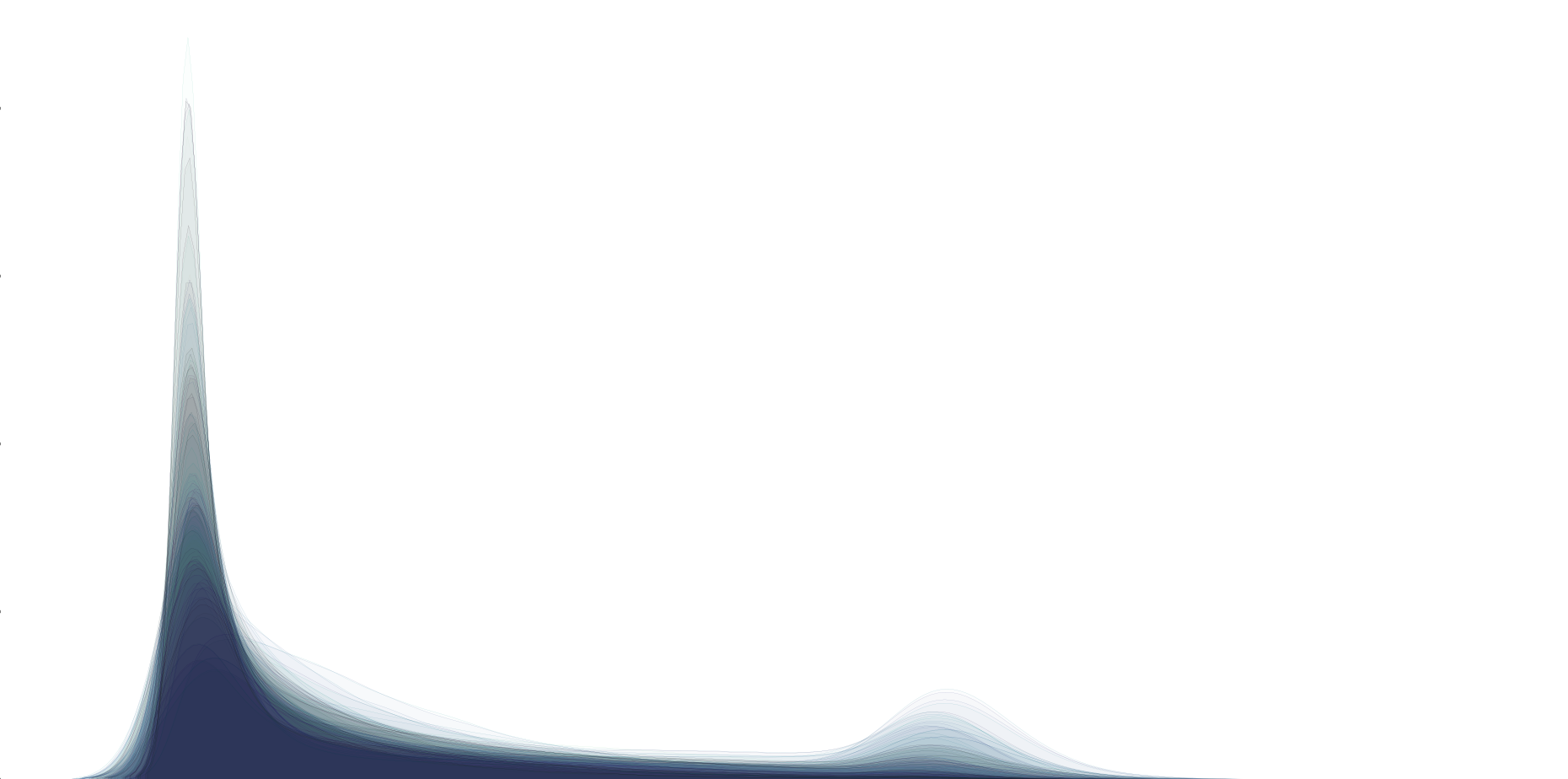}}                                 \\ \hline
\end{tabular}
}
%\vspace*{-3mm}
\end{table}

\section{Bang for the buck}\label{sec:eval}

\begin{figure*}[t!]
\centering
\includegraphics[width=1\linewidth]{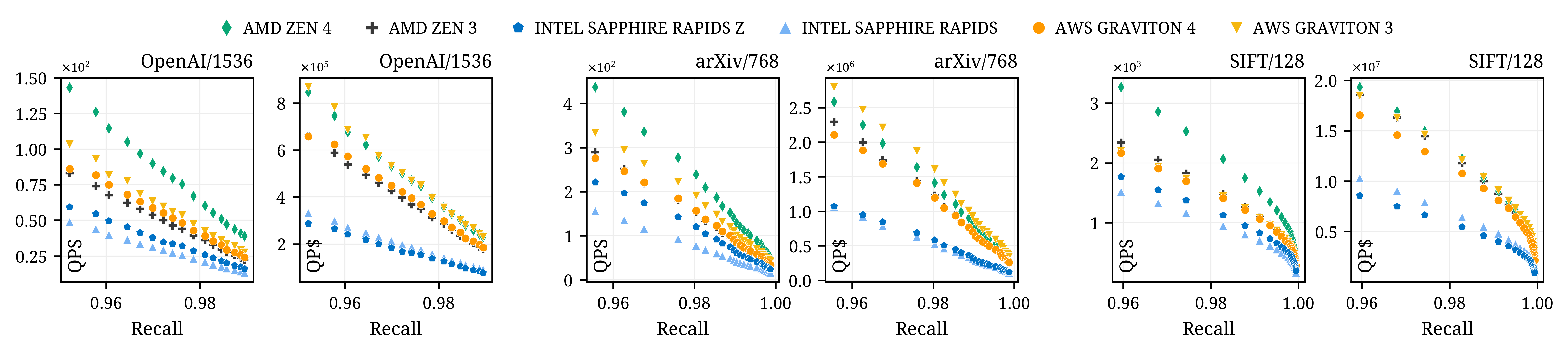}
\vspace*{-8mm}
\caption{QPS and QP\$ of cloud instances running queries on a FAISS IVF index without quantization (float32). The Zen4 microarchitecture takes the upper hand in QPS, but Graviton3 takes a slight advantage in QP\$.}
\label{fig:ivf}
\vspace*{-2mm}
\end{figure*}

We benchmarked the end-to-end search latency of HNSW, IVF indexes, and \textit{full scans} (i.e., without an index), with vectors quantized at different levels on five microarchitectures available in AWS. These are presented in Table ~\ref{tab:machine_details} alongside their on-demand price in {\small\tt us-east-1} at the time of doing this study. These cover the major ISAs and popular CPUs. For Intel SPR, we also present benchmarks on the $Z$ series variant\footnote{\href{https://aws.amazon.com/ec2/instance-types/r7iz/}{aws.amazon.com: r7iz instances}}. All machines have 64GB of DRAM, 8 vCPUs, and Ubuntu 24.04 as OS. For each experiment, we report queries-per-second (QPS) and queries-per-dollar (QP\$). % Note that we have selected instances that share the same configurations of vCPUs and RAM.  

For our experiments, we used FAISS (v1.9.0)~\cite{faisscode, faisspaper} compiled to target the underlying CPU capabilities. We chose FAISS as it is the cornerstone of many vector systems. For instance, Milvus~\cite{milvus} and Weaviate~\cite{weaviate} vector engines started as a fork of FAISS. OpenSearch (AWS) is directly built on top of FAISS\footnote{\href{https://aws.amazon.com/blogs/big-data/amazon-opensearch-services-vector-database-capabilities-explained/}{aws.amazon.com: Amazon OpenSearch Service’s VecDB capabilities explained}}. 
By using FAISS, we also avoid introducing possible artifacts of vector databases (e.g., Milvus \textit{dynamic batching} mechanism that executes queries at intervals). 

\vfill
\pagebreak

Unfortunately, when using SQ, {\small \tt float16}, or {\small \tt bfloat}, FAISS first decodes vectors back to the {\small \tt float32} domain to perform the distance calculations. To make up for that, we add USearch (v.2.16.9)~\cite{usearch} to our benchmarks. USearch is a vector engine focused on getting the best performance on different CPUs with symmetric kernels for various distance metrics and quantized types on HNSW indexes and full scans. USearch is currently used by ClickHouse and DuckDB on their VSS extensions. %Note that the purpose of our study is not to compare the performance of these libraries. % For example USearch is known to only perform better than FAISS in the billion-vectors use case.

For our analysis, we have chosen three datasets that exhibit different dimensionalities, presented in Table~\ref{tab:datasets}. These datasets are commonly used to evaluate VSS techniques~\cite{annbench, candy}. From these collections, one represents vectors from image data (SIFT/128), and two represent vector embeddings from text (arXiv/768, OpenAI/1536). % We perform our experiments of IVF indexes in FAISS and HNSW indexes in USearch. % Note that the purpose of our study is not to compare the performance of these libraries.

\vspace*{3mm}
\noindent{\bf Indexes hyperparameters. } \underline{FAISS IVF}: Number of centroids: $4*\sqrt{n}$, buckets visited (for recall tuning): from 2 to 512, training points: all. Quantized with 8, 6, 4, and 1-bit. \underline{FAISS FULL SCAN}: Quantized with 8, 6, 4, 1-bit and PQ ({\small \tt n\_bits}: 8, {\small \tt subspaces m}: $\frac{D}{4}$). \underline{USearch HNSW}: {\small \tt m}: 16, {\small \tt ef\_construction}: 128, {\small \tt ef\_search} (for recall tuning): from 2 to 512. Quantized with {\small \tt float16}, {\small \tt bfloat}, 8, and 1-bit. \underline{USearch FULL SCAN}: Quantized with {\small \tt float16}, {\small \tt bfloat}, 8, and 1-bit. In all settings, we ran queries individually (no multi-threading) with $k=10$, using L2 and Hamming (on BQ) as distance metrics.

\begin{table}[t!]
\renewcommand{\tabcolsep}{1.5pt}
\centering
\caption{QPS and QP\$ of cloud instances running queries on FAISS IVF indexes quantized at different levels. 
%Zen4 is the overall winner, giving > 2x more QP\$ than Intel on float32 vectors. However, the difference gap closes with the decrease of dimensionality. Graviton 4 is outperformed in quantized vectors but takes the upper hand in binary vectors. 
\textit{QP\$ values are in the order of 10\textsuperscript{5}. Color coding is the same as in Table~\ref{tab:summary}.} %\colorbox[HTML]{fefbcb}{Yellow} and \colorbox[HTML]{ffbfbe}{red} indicate 2x and 3x less relative performance to the best, respectively.}
}
\vspace*{-4mm}
\label{tab:ivf}
\resizebox{1\linewidth}{!}{%
\begin{tabular}{l|l|cc|cc|cc|cc|cc|}
\cline{2-12}
                                                    & \multicolumn{1}{c|}{}                                      & \multicolumn{2}{c|}{\textbf{float32}}                  & \multicolumn{2}{c|}{\textbf{SQ8 (8-bit)}}              & \multicolumn{2}{c|}{\textbf{SQ6 (6-bit)}}              & \multicolumn{2}{c|}{\textbf{SQ4 (4-bit)}}              & \multicolumn{2}{c|}{\textbf{BQ (1-bit)}}                 \\
                                                    & \multicolumn{1}{c|}{\multirow{-2}{*}{\textbf{Microarch.}}} & QPS            & /\$                                   & QPS            & /\$                                   & QPS            & /\$                                   & QPS            & /\$                                   & QPS             & /\$                                    \\ \hline
\multicolumn{1}{|l|}{}                              & Graviton3                                                  & 27.4           & \cellcolor[HTML]{A6FFA6}\textbf{2.3}  & 9.6            & \cellcolor[HTML]{FFBFBE}0.8           & 5.8            & \cellcolor[HTML]{FFBFBE}0.5           & 6.7            & \cellcolor[HTML]{FFBFBE}0.6           & 504.3           & \cellcolor[HTML]{A6FFA6}\textbf{42.4}  \\
\multicolumn{1}{|l|}{}                              & Zen3                                                       & 22.4           & 1.8                                   & 32.5           & 2.6                                   & 15.0           & 1.2                                   & 17.7           & 1.4                                   & 363.8           & 28.9                                   \\
\multicolumn{1}{|l|}{}                              & Graviton4                                                  & 24.2           & 1.8                                   & 11.7           & \cellcolor[HTML]{FFBFBE}0.9           & 7.3            & \cellcolor[HTML]{FFBFBE}0.6           & 8.1            & \cellcolor[HTML]{FFBFBE}0.6           & \textbf{529.1}  & \cellcolor[HTML]{E9FFE9}40.4           \\
\multicolumn{1}{|l|}{}                              & SPR                                                        & 13.3           & \cellcolor[HTML]{FEFBCB}0.9           & 28.9           & 2.0                                   & 30.9           & \cellcolor[HTML]{E9FFE9}2.1           & 30.9           & \cellcolor[HTML]{A6FFA6}\textbf{2.1}  & 421.1           & 28.6                                   \\
\multicolumn{1}{|l|}{}                              & Zen4                                                       & \textbf{38.9}  & \cellcolor[HTML]{A6FFA6}\textbf{2.3}  & \textbf{57.3}  & \cellcolor[HTML]{A6FFA6}\textbf{3.4}  & \textbf{39.3}  & \cellcolor[HTML]{A6FFA6}\textbf{2.3}  & \textbf{36.0}  & \cellcolor[HTML]{A6FFA6}\textbf{2.1}  & 389.4           & 23                                     \\
\multicolumn{1}{|l|}{\multirow{-6}{*}{\rotatebox[origin=c]{90}{\raisebox{0.3\normalbaselineskip}[0pt][0pt]{\small OpenAI/1536}}}} & SPR Z                                                      & 16.1           & \cellcolor[HTML]{FEFBCB}0.8           & 31.0           & \cellcolor[HTML]{FEFBCB}1.5           & 34.7           & 1.7                                   & 32.1           & 1.6                                   & 469.3           & 22.7                                   \\ \hline
\multicolumn{1}{|l|}{}                              & Graviton3                                                  & 43.0           & \cellcolor[HTML]{A6FFA6}\textbf{3.6}  & 13.6           & \cellcolor[HTML]{FFBFBE}1.1           & 9.4            & \cellcolor[HTML]{FFBFBE}0.8           & 8.8            & \cellcolor[HTML]{FFBFBE}0.7           & 993.0           & \cellcolor[HTML]{E9FFE9}83.4           \\
\multicolumn{1}{|l|}{}                              & Zen3                                                       & 32.5           & 2.6                                   & 45.9           & \cellcolor[HTML]{E9FFE9}3.6           & 24.2           & 1.9                                   & 23.4           & 1.9                                   & 847.5           & 67.3                                   \\
\multicolumn{1}{|l|}{}                              & Graviton4                                                  & 36.2           & 2.8                                   & 16.7           & \cellcolor[HTML]{FFBFBE}1.3           & 11.8           & \cellcolor[HTML]{FFBFBE}0.9           & 10.7           & \cellcolor[HTML]{FFBFBE}0.8           & \textbf{1204.8} & \cellcolor[HTML]{A6FFA6}\textbf{92.1}  \\
\multicolumn{1}{|l|}{}                              & SPR                                                        & 17.6           & \cellcolor[HTML]{FFBFBE}1.2           & 40.1           & 2.7                                   & 49.5           & \cellcolor[HTML]{A6FFA6}\textbf{3.4}  & 38.0           & \cellcolor[HTML]{E9FFE9}2.6           & 612.4           & \cellcolor[HTML]{FEFBCB}41.7           \\
\multicolumn{1}{|l|}{}                              & Zen4                                                       & \textbf{54.2}  & \cellcolor[HTML]{E9FFE9}3.2           & \textbf{67.9}  & \cellcolor[HTML]{A6FFA6}\textbf{4.0}  & 55.7           & \cellcolor[HTML]{E9FFE9}3.3           & \textbf{45.2}  & \cellcolor[HTML]{A6FFA6}\textbf{2.7}  & 840.3           & 49.7                                   \\
\multicolumn{1}{|l|}{\multirow{-6}{*}{\rotatebox[origin=c]{90}{\raisebox{0.3\normalbaselineskip}[0pt][0pt]{\small arXiv/768}}}}   & SPR Z                                                      & 25.4           & \cellcolor[HTML]{FFBFBE}1.2           & 46.2           & 2.2                                   & \textbf{61.6}  & \cellcolor[HTML]{E9FFE9}3.0           & 45.1           & \cellcolor[HTML]{E9FFE9}2.2           & 684.9           & \cellcolor[HTML]{FEFBCB}33.1           \\ \hline
\multicolumn{1}{|l|}{}                              & Graviton3                                                  & 557.1          & \cellcolor[HTML]{A6FFA6}\textbf{46.8} & 206.2          & \cellcolor[HTML]{FFBFBE}17.3          & 157.1          & \cellcolor[HTML]{FFBFBE}13.2          & 178.0          & \cellcolor[HTML]{FFBFBE}15.0          & 5681.8          & \cellcolor[HTML]{E9FFE9}476.6          \\
\multicolumn{1}{|l|}{}                              & Zen3                                                       & 540.0          & \cellcolor[HTML]{E9FFE9}42.9          & 612.0          & \cellcolor[HTML]{E9FFE9}48.6          & 384.8          & 30.5                                  & 439.6          & 34.9                                  & \textbf{6250.0} & \cellcolor[HTML]{A6FFA6}\textbf{497.5} \\
\multicolumn{1}{|l|}{}                              & Graviton4                                                  & 538.2          & \cellcolor[HTML]{E9FFE9}41.1          & 249.9          & \cellcolor[HTML]{FEFBCB}19.1          & 192.0          & \cellcolor[HTML]{FFBFBE}14.7          & 212.6          & \cellcolor[HTML]{FFBFBE}16.2          & \textbf{6250.0} & \cellcolor[HTML]{E9FFE9}476.6          \\
\multicolumn{1}{|l|}{}                              & SPR                                                        & 342.1          & \cellcolor[HTML]{FEFBCB}23.3          & 749.6          & \cellcolor[HTML]{E9FFE9}51.0          & 871.1          & \cellcolor[HTML]{A6FFA6}\textbf{59.3} & \textbf{831.3} & \cellcolor[HTML]{A6FFA6}\textbf{56.5} & 5181.3          & 351.7                                  \\
\multicolumn{1}{|l|}{}                              & Zen4                                                       & \textbf{745.2} & \cellcolor[HTML]{E9FFE9}44.1          & \textbf{886.5} & \cellcolor[HTML]{A6FFA6}\textbf{52.4} & 849.6          & \cellcolor[HTML]{E9FFE9}50.2          & 814.3          & \cellcolor[HTML]{E9FFE9}48.2          & 5291.0          & 313.1                                  \\
\multicolumn{1}{|l|}{\multirow{-6}{*}{\rotatebox[origin=c]{90}{\raisebox{0.3\normalbaselineskip}[0pt][0pt]{\small SIFT/128}}}}    & SPR Z                                                      & 386.0          & \cellcolor[HTML]{FEFBCB}18.7          & 833.3          & 40.3                                  & \textbf{898.5} & 43.5                                  & 821.0          & 39.7                                  & 5681.8          & 274.4                                  \\ \hline
\end{tabular}
}
\vspace*{-4mm}
\end{table}
\begin{table}[t!]
\renewcommand{\tabcolsep}{1.5pt}
\centering
\caption{QPS and QP\$ of cloud instances running queries on USearch HNSW quantized at different levels. 
%Intel Z always yield the highest QPS. However, Zen3 and Graviton give more QP\$ in float32 and bfloat. Intel takes the upper hand on float16, SQ8 and BQ. 
\textit{QP\$ values are in the order of 10\textsuperscript{5}. Color coding is the same as in Table~\ref{tab:summary}.} %\colorbox[HTML]{fefbcb}{Yellow} and \colorbox[HTML]{ffbfbe}{red} cells indicate 2x and 3x less relative performance to the best, respectively.
}
\vspace*{-4mm}
\label{tab:hnsw}
\resizebox{1\columnwidth}{!}{%
\begin{tabular}{l|l|cc|cc|cc|cc|cc|}
\cline{2-12}
                                                    & \multicolumn{1}{c|}{}                                      & \multicolumn{2}{c|}{\textbf{float32}}                 & \multicolumn{2}{c|}{\textbf{float16}}                   & \multicolumn{2}{c|}{\textbf{bfloat}}                   & \multicolumn{2}{c|}{\textbf{SQ8 (8-bit)}}              & \multicolumn{2}{c|}{\textbf{BQ (1-bit)}}                \\  
                                                    & \multicolumn{1}{c|}{\multirow{-2}{*}{\textbf{Microarch.}}} & QPS            & /\$                                  & QPS             & /\$                                   & QPS             & /\$                                  & QPS             & /\$                                  & QPS             & /\$                                   \\ \hline
\multicolumn{1}{|l|}{}                              & Graviton3                                                  & 192.0          & \cellcolor[HTML]{A6FFA6}\textbf{1.6} & 247.3           & \cellcolor[HTML]{A6FFA6}\textbf{2.1}  & 196.3           & \cellcolor[HTML]{A6FFA6}\textbf{1.6} & 345.3           & \cellcolor[HTML]{A6FFA6}\textbf{2.9} & 639.0           & 5.4                                   \\
\multicolumn{1}{|l|}{}                              & Zen3                                                       & 194.2          & \cellcolor[HTML]{E9FFE9}1.5          & 197.1           & 1.6                                   & 191.7           & \cellcolor[HTML]{E9FFE9}1.5          & 265.0           & 2.1                                  & 462.5           & \cellcolor[HTML]{FEFBCB}3.7           \\
\multicolumn{1}{|l|}{}                              & Graviton4                                                  & 183.6          & \cellcolor[HTML]{E9FFE9}1.4          & 241.4           & \cellcolor[HTML]{E9FFE9}1.8           & 204.0           & \cellcolor[HTML]{A6FFA6}\textbf{1.6} & 310.9           & \cellcolor[HTML]{E9FFE9}2.4          & 620.9           & 4.7                                   \\
\multicolumn{1}{|l|}{}                              & SPR                                                        & 182.0          & 1.2                                  & 270.2           & \cellcolor[HTML]{E9FFE9}1.8           & 198.6           & \cellcolor[HTML]{E9FFE9}1.4          & 331.2           & 2.3                                  & 1140.4          & \cellcolor[HTML]{A6FFA6}\textbf{7.8}  \\
\multicolumn{1}{|l|}{}                              & Zen4                                                       & 165.6          & 1.0                                  & 168.0           & \cellcolor[HTML]{FEFBCB}1.0           & 194.9           & 1.2                                  & 279.5           & 1.7                                  & 838.8           & 5.0                                   \\
\multicolumn{1}{|l|}{\multirow{-6}{*}{\rotatebox[origin=c]{90}{\raisebox{0.3\normalbaselineskip}[0pt][0pt]{\small OpenAI/1536}}}} & SPR Z                                                      & \textbf{228.7} & 1.1                                  & \textbf{359.3}  & \cellcolor[HTML]{E9FFE9}1.7           & \textbf{235.9}  & 1.1                                  & \textbf{389.3}  & 1.9                                  & \textbf{1313.9} & \cellcolor[HTML]{E9FFE9}6.4           \\ \hline
\multicolumn{1}{|l|}{}                              & Graviton3                                                  & 212.6          & \cellcolor[HTML]{E9FFE9}1.8          & 279.8           & \cellcolor[HTML]{E9FFE9}2.4           & 226.9           & \cellcolor[HTML]{A6FFA6}\textbf{1.9} & 327.9           & \cellcolor[HTML]{E9FFE9}2.8          & 554.9           & 4.7                                   \\
\multicolumn{1}{|l|}{}                              & Zen3                                                       & 242.4          & \cellcolor[HTML]{A6FFA6}\textbf{1.9} & 237.9           & 1.9                                   & 229.8           & \cellcolor[HTML]{E9FFE9}1.8          & 319.7           & \cellcolor[HTML]{E9FFE9}2.5          & 585.0           & 4.6                                   \\
\multicolumn{1}{|l|}{}                              & Graviton4                                                  & 224.0          & \cellcolor[HTML]{E9FFE9}1.7          & 276.5           & \cellcolor[HTML]{E9FFE9}2.1           & 227.3           & \cellcolor[HTML]{E9FFE9}1.7          & 317.0           & 2.4                                  & 664.1           & 5.1                                   \\
\multicolumn{1}{|l|}{}                              & SPR                                                        & 229.3          & \cellcolor[HTML]{E9FFE9}1.6          & 382.4           & \cellcolor[HTML]{A6FFA6}\textbf{2.6}  & 238.4           & \cellcolor[HTML]{E9FFE9}1.6          & \textbf{450.3}  & \cellcolor[HTML]{A6FFA6}\textbf{3.1} & 965.3           & \cellcolor[HTML]{A6FFA6}\textbf{6.6}  \\
\multicolumn{1}{|l|}{}                              & Zen4                                                       & 212.3          & 1.3                                  & 205.0           & \cellcolor[HTML]{FEFBCB}1.2           & 244.8           & 1.4                                  & 307.8           & 1.8                                  & 718.4           & 4.2                                   \\
\multicolumn{1}{|l|}{\multirow{-6}{*}{\rotatebox[origin=c]{90}{\raisebox{0.3\normalbaselineskip}[0pt][0pt]{\small arXiv/768}}}}   & SPR Z                                                      & \textbf{297.7} & 1.4                                  & \textbf{472.7}  & \cellcolor[HTML]{E9FFE9}2.3           & \textbf{254.7}  & 1.2                                  & 432.7           & 2.1                                  & \textbf{965.9}  & 4.7                                   \\ \hline
\multicolumn{1}{|l|}{}                              & Graviton3                                                  & 626.7          & \cellcolor[HTML]{E9FFE9}5.3          & 1299.2          & 10.9                                  & 479.4           & 4.0                                  & 870.9           & \cellcolor[HTML]{E9FFE9}7.3          & 1527.2          & 12.8                                  \\
\multicolumn{1}{|l|}{}                              & Zen3                                                       & 693.4          & \cellcolor[HTML]{E9FFE9}5.5          & 517.5           & \cellcolor[HTML]{FFBFBE}4.1           & 505.7           & 4.0                                  & 812.5           & 6.4                                  & 1879.7          & 14.9                                  \\
\multicolumn{1}{|l|}{}                              & Graviton4                                                  & 555.0          & 4.2                                  & 1583.5          & 12.1                                  & 437.5           & 3.3                                  & 1024.4          & \cellcolor[HTML]{E9FFE9}7.8          & 1600.8          & 12.2                                  \\
\multicolumn{1}{|l|}{}                              & SPR                                                        & 905.8          & \cellcolor[HTML]{A6FFA6}\textbf{6.2} & 2337.0          & \cellcolor[HTML]{A6FFA6}\textbf{15.9} & 918.7           & \cellcolor[HTML]{A6FFA6}\textbf{6.2} & 1338.3          & \cellcolor[HTML]{A6FFA6}\textbf{9.1} & 2907.0          & \cellcolor[HTML]{A6FFA6}\textbf{19.8} \\
\multicolumn{1}{|l|}{}                              & Zen4                                                       & 704.7          & 4.2                                  & 616.4           & \cellcolor[HTML]{FFBFBE}3.6           & 773.5           & 4.6                                  & 901.5           & 5.3                                  & 2136.8          & 12.6                                  \\
\multicolumn{1}{|l|}{\multirow{-6}{*}{\rotatebox[origin=c]{90}{\raisebox{0.3\normalbaselineskip}[0pt][0pt]{\small SIFT/128}}}}    & SPR Z                                                      & \textbf{959.7} & 4.6                                  & \textbf{2866.2} & \cellcolor[HTML]{E9FFE9}13.9          & \textbf{1071.6} & \cellcolor[HTML]{E9FFE9}5.2          & \textbf{1542.3} & \cellcolor[HTML]{E9FFE9}7.5          & \textbf{3510.0} & \cellcolor[HTML]{E9FFE9}17.0          \\ \hline
\end{tabular}
}
\vspace*{-4mm}
\end{table}

\subsection{IVF}\label{sec:eval-ivf}
Figure~\ref{fig:ivf} shows the performance of the microarchitectures on an IVF index without quantization at different recall levels. In this setting, Zen4 is the clear winner across all datasets in QPS, giving 3x the performance of both Intels. However, in QP\$, Graviton3 ties with Zen4. Zen3 and Graviton4 follow closely in QP\$, and their gap closes as dimensionality decreases. Another observation is that the targeted recall does not affect the relative performance of microarchitectures. 

Table~\ref{tab:ivf} shows the performance of each microarchitecture with different quantization levels at the highest possible recall. In SQ vectors, Zen4 and Intel perform well. However, Zen4 takes the upper hand in higher dimensionalities. The Gravitons heavily underperform in SQ vectors (up to 3x and 6x less QP\$ and QPS, resp.). This is because the asymmetric distance calculation in FAISS requires going from the quantized domain to the {\small \tt float32} domain. % The latter can be done with {\small \tt HI/LO-UNPACK} or {\small \tt CAST} on 8 and 4-bit and {\small \tt SHIFT+OR} or {\small \tt SHUFFLE} on 6-bits. 
The latter can be done with SIMD instructions (e.g., {\small \tt HI/LO-UNPACK} on 8-bit and {\small \tt SHIFT+OR} on 6-bit vectors). 
However, FAISS currently uses scalar code to do this in ARM CPUs, contrary to the fully SIMDized kernels used in AVX2/AVX512. Nevertheless, the Gravitons shine on 1-bit vectors as FAISS \textit{does} implement a specialized kernel for BQ in NEON. The latter shows the importance of SIMD kernels. % Interestingly, the {\small \tt POPCOUNT} instruction in Sapphire Rapids and Zen4 used in FAISS is not on par with the other microarchitectures for 1-bit vectors.

\begin{figure*}[t!]
\centering
\includegraphics[width=1\linewidth]{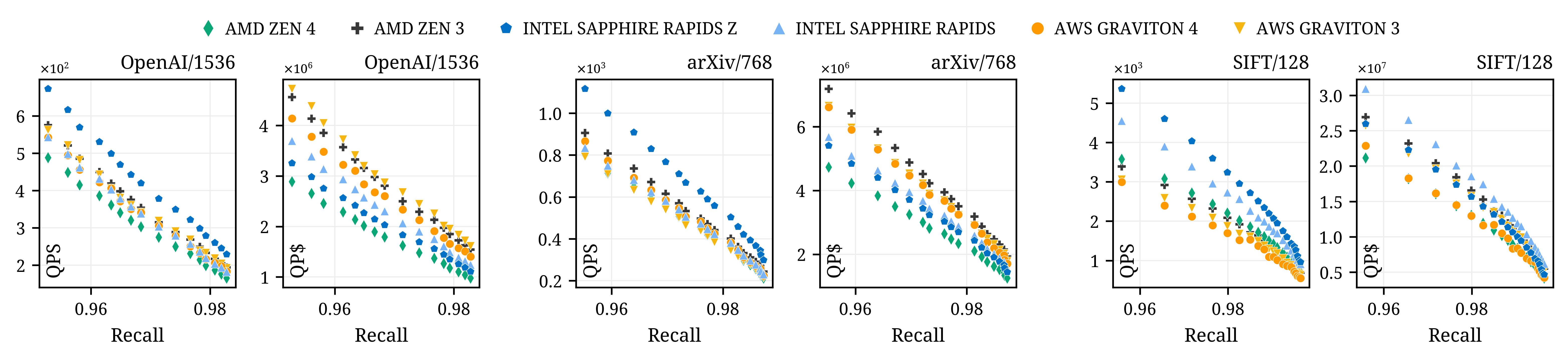}
\vspace*{-8mm}
\caption{QPS and QP\$ of cloud instances running queries on a flat HNSW index (float32). Intel Z takes the upper hand in QPS, but Zen3 and Graviton3 give higher QP\$ on high-dimensional vectors. Here, contrary to IVF indexes, Zen4 underperforms.}
\label{fig:hnsw}
\vspace*{-2mm}
\end{figure*}

\begin{table*}[t]
\renewcommand{\tabcolsep}{2.5pt}
\centering
\caption{QPS and QP\$ of cloud instances running full scan queries. Intel performs the best in PQ vectors. The Gravitons take the upper hand in float16 and 1-bit vectors. In all the other settings, Zen4 is the overall winner. The Gravitons perform well in USearch SQ but not in FAISS SQ due to the absence of SIMD. 
\textit{QP\$ is in the order of 10\textsuperscript{4}. Color coding is the same as Table~\ref{tab:summary}.}
}
\vspace*{-4mm}
\label{tab:full}
\resizebox{1\linewidth}{!}{%
\begin{tabular}{ll|ccccccccccrr|cccccccccc|}
\cline{3-24}
                                                     &                                                            & \multicolumn{12}{c|}{\textbf{FAISS FULL SCAN}}                                                                                                                                                                                                                                                                                                                                                                                                                      & \multicolumn{10}{c|}{\textbf{USearch FULL SCAN}}                                                                                                                                                                                                                                                                                                                          \\ \cline{2-24} 
\multicolumn{1}{l|}{}                                & \multicolumn{1}{c|}{}                                      & \multicolumn{2}{c|}{\textbf{float32}}                                      & \multicolumn{2}{c|}{\textbf{SQ8 (8-bit)}}                                  & \multicolumn{2}{c|}{\textbf{SQ6 (6-bit)}}                                  & \multicolumn{2}{c|}{\textbf{SQ4 (4-bit)}}                                  & \multicolumn{2}{c|}{\textbf{BQ (1-bit)}}                                      & \multicolumn{2}{c|}{\textbf{PQ}}                                & \multicolumn{2}{c|}{\textbf{float32}}                                      & \multicolumn{2}{c|}{\textbf{float16}}                                      & \multicolumn{2}{c|}{\textbf{bfloat}}                                       & \multicolumn{2}{c|}{\textbf{SQ8 (8-bit)}}                                  & \multicolumn{2}{c|}{\textbf{BQ (1-bit)}}              \\
\multicolumn{1}{l|}{}                                & \multicolumn{1}{c|}{\multirow{-2}{*}{\textbf{Microarch.}}} & QPS           & \multicolumn{1}{c|}{/\$}                                   & QPS           & \multicolumn{1}{c|}{/\$}                                   & QPS           & \multicolumn{1}{c|}{/\$}                                   & QPS           & \multicolumn{1}{c|}{/\$}                                   & QPS             & \multicolumn{1}{c|}{/\$}                                    & \multicolumn{1}{c}{QPS} & \multicolumn{1}{c|}{/\$}              & QPS           & \multicolumn{1}{c|}{/\$}                                   & QPS           & \multicolumn{1}{c|}{/\$}                                   & QPS           & \multicolumn{1}{c|}{/\$}                                   & QPS           & \multicolumn{1}{c|}{/\$}                                   & QPS           & /\$                                   \\ \hline
\multicolumn{1}{|l|}{}                               & Graviton3                                                  & 3.9           & \multicolumn{1}{c|}{\cellcolor[HTML]{E9FFE9}3.2}           & 1.2           & \multicolumn{1}{c|}{\cellcolor[HTML]{FFBFBE}1.0}           & 0.8           & \multicolumn{1}{c|}{\cellcolor[HTML]{FFBFBE}0.7}           & 0.9           & \multicolumn{1}{c|}{\cellcolor[HTML]{FFBFBE}0.8}           & 82.5            & \multicolumn{1}{c|}{\cellcolor[HTML]{A6FFA6}\textbf{69.4}}  & 4.1                     & 3.5                                   & 3.7           & \multicolumn{1}{c|}{\cellcolor[HTML]{A6FFA6}\textbf{3.1}}  & \textbf{6.8}  & \multicolumn{1}{c|}{\cellcolor[HTML]{A6FFA6}\textbf{5.8}}  & 2.9           & \multicolumn{1}{c|}{\cellcolor[HTML]{FEFBCB}2.5}           & 12.8          & \multicolumn{1}{c|}{\cellcolor[HTML]{A6FFA6}\textbf{10.8}} & 37.5          & \cellcolor[HTML]{E9FFE9}31.5          \\
\multicolumn{1}{|l|}{}                               & Zen3                                                       & 3.6           & \multicolumn{1}{c|}{\cellcolor[HTML]{E9FFE9}2.8}           & 4.4           & \multicolumn{1}{c|}{\cellcolor[HTML]{A6FFA6}\textbf{3.5}}  & 2.1           & \multicolumn{1}{c|}{1.6}                                   & 2.5           & \multicolumn{1}{c|}{2.0}                                   & 62.7            & \multicolumn{1}{c|}{49.8}                                   & 4.2                     & 3.4                                   & 3.3           & \multicolumn{1}{c|}{\cellcolor[HTML]{E9FFE9}2.6}           & 4.0           & \multicolumn{1}{c|}{3.2}                                   & 3.9           & \multicolumn{1}{c|}{3.1}                                   & 7.6           & \multicolumn{1}{c|}{6.1}                                   & 28.9          & 22.9                                  \\
\multicolumn{1}{|l|}{}                               & Graviton4                                                  & 3.6           & \multicolumn{1}{c|}{\cellcolor[HTML]{E9FFE9}2.7}           & 1.5           & \multicolumn{1}{c|}{\cellcolor[HTML]{FEFBCB}1.2}           & 1.0           & \multicolumn{1}{c|}{\cellcolor[HTML]{FFBFBE}0.8}           & 1.1           & \multicolumn{1}{c|}{\cellcolor[HTML]{FFBFBE}0.8}           & \textbf{90.3}   & \multicolumn{1}{c|}{\cellcolor[HTML]{E9FFE9}69}             & 4.3                     & 3.3                                   & 3.3           & \multicolumn{1}{c|}{\cellcolor[HTML]{E9FFE9}2.5}           & 6.2           & \multicolumn{1}{c|}{\cellcolor[HTML]{E9FFE9}4.7}           & 3.6           & \multicolumn{1}{c|}{2.8}                                   & 13.9          & \multicolumn{1}{c|}{\cellcolor[HTML]{E9FFE9}10.6}          & 46.2          & \cellcolor[HTML]{A6FFA6}\textbf{35.3} \\
\multicolumn{1}{|l|}{}                               & SPR                                                        & 1.9           & \multicolumn{1}{c|}{\cellcolor[HTML]{FEFBCB}1.3}           & 4.0           & \multicolumn{1}{c|}{2.7}                                   & 4.6           & \multicolumn{1}{c|}{\cellcolor[HTML]{A6FFA6}\textbf{3.1}}  & 4.7           & \multicolumn{1}{c|}{\cellcolor[HTML]{A6FFA6}\textbf{3.2}}  & 49.2            & \multicolumn{1}{c|}{\cellcolor[HTML]{FEFBCB}33.5}           & 6.6                     & \cellcolor[HTML]{A6FFA6}\textbf{4.5}  & 1.9           & \multicolumn{1}{c|}{\cellcolor[HTML]{FEFBCB}1.3}           & 3.7           & \multicolumn{1}{c|}{\cellcolor[HTML]{FEFBCB}2.5}           & 2.9           & \multicolumn{1}{c|}{\cellcolor[HTML]{FEFBCB}2.0}           & 6.1           & \multicolumn{1}{c|}{\cellcolor[HTML]{FEFBCB}4.1}           & 34.0          & 23.1                                  \\
\multicolumn{1}{|l|}{}                               & Zen4                                                       & \textbf{5.6}  & \multicolumn{1}{c|}{\cellcolor[HTML]{A6FFA6}\textbf{3.3}}  & \textbf{6.0}  & \multicolumn{1}{c|}{\cellcolor[HTML]{A6FFA6}\textbf{3.5}}  & 4.7           & \multicolumn{1}{c|}{\cellcolor[HTML]{E9FFE9}2.8}           & \textbf{4.9}  & \multicolumn{1}{c|}{\cellcolor[HTML]{E9FFE9}2.9}           & 77.5            & \multicolumn{1}{c|}{45.8}                                   & 4.0                     & 2.4                                   & \textbf{5.2}  & \multicolumn{1}{c|}{\cellcolor[HTML]{A6FFA6}\textbf{3.1}}  & 4.3           & \multicolumn{1}{c|}{\cellcolor[HTML]{FEFBCB}2.6}           & \textbf{8.8}  & \multicolumn{1}{c|}{\cellcolor[HTML]{A6FFA6}\textbf{5.2}}  & \textbf{14.3} & \multicolumn{1}{c|}{8.4}                                   & \textbf{46.7} & 27.6                                  \\
\multicolumn{1}{|l|}{\multirow{-6}{*}{\rotatebox[origin=c]{90}{\raisebox{-0.25\normalbaselineskip}[2pt][2pt]{\small OpenAI/1536}}}} & SPR Z                                                      & 2.3           & \multicolumn{1}{c|}{\cellcolor[HTML]{FFBFBE}1.1}           & 4.2           & \multicolumn{1}{c|}{2.0}                                   & \textbf{4.8}  & \multicolumn{1}{c|}{2.3}                                   & \textbf{4.9}  & \multicolumn{1}{c|}{2.4}                                   & 57.1            & \multicolumn{1}{c|}{\cellcolor[HTML]{FEFBCB}27.6}           & \textbf{7.0}            & 3.4                                   & 2.3           & \multicolumn{1}{c|}{\cellcolor[HTML]{FEFBCB}1.1}           & 4.6           & \multicolumn{1}{c|}{\cellcolor[HTML]{FEFBCB}2.2}           & 3.1           & \multicolumn{1}{c|}{\cellcolor[HTML]{FFBFBE}1.5}           & 6.8           & \multicolumn{1}{c|}{\cellcolor[HTML]{FFBFBE}3.3}           & 40.8          & 19.7                                  \\ \hline
\multicolumn{1}{|l|}{}                               & Graviton3                                                  & 3.4           & \multicolumn{1}{c|}{\cellcolor[HTML]{E9FFE9}2.8}           & 1.1           & \multicolumn{1}{c|}{\cellcolor[HTML]{FFBFBE}0.9}           & 0.7           & \multicolumn{1}{c|}{\cellcolor[HTML]{FFBFBE}0.6}           & 0.8           & \multicolumn{1}{c|}{\cellcolor[HTML]{FFBFBE}0.7}           & \textbf{71.4}   & \multicolumn{1}{c|}{\cellcolor[HTML]{A6FFA6}\textbf{60}}    & 4.2                     & \cellcolor[HTML]{E9FFE9}3.5           & 3.4           & \multicolumn{1}{c|}{\cellcolor[HTML]{A6FFA6}\textbf{2.9}}  & \textbf{6.0}  & \multicolumn{1}{c|}{\cellcolor[HTML]{A6FFA6}\textbf{5.0}}  & 3.3           & \multicolumn{1}{c|}{2.8}                                   & 9.7           & \multicolumn{1}{c|}{\cellcolor[HTML]{E9FFE9}8.2}           & 20.5          & \cellcolor[HTML]{E9FFE9}17.3          \\
\multicolumn{1}{|l|}{}                               & Zen3                                                       & 3.0           & \multicolumn{1}{c|}{\cellcolor[HTML]{E9FFE9}2.4}           & 3.8           & \multicolumn{1}{c|}{\cellcolor[HTML]{E9FFE9}3.0}           & 1.8           & \multicolumn{1}{c|}{1.5}                                   & 2.2           & \multicolumn{1}{c|}{1.7}                                   & 48.7            & \multicolumn{1}{c|}{38.6}                                   & 4.5                     & \cellcolor[HTML]{E9FFE9}3.6           & 2.7           & \multicolumn{1}{c|}{2.2}                                   & 3.3           & \multicolumn{1}{c|}{2.6}                                   & 3.2           & \multicolumn{1}{c|}{2.5}                                   & 5.8           & \multicolumn{1}{c|}{4.6}                                   & 19.0          & \cellcolor[HTML]{E9FFE9}15.1          \\
\multicolumn{1}{|l|}{}                               & Graviton4                                                  & 3.3           & \multicolumn{1}{c|}{\cellcolor[HTML]{E9FFE9}2.5}           & 1.3           & \multicolumn{1}{c|}{\cellcolor[HTML]{FFBFBE}1.0}           & 0.9           & \multicolumn{1}{c|}{\cellcolor[HTML]{FFBFBE}0.7}           & 1.0           & \multicolumn{1}{c|}{\cellcolor[HTML]{FFBFBE}0.7}           & 66.6            & \multicolumn{1}{c|}{\cellcolor[HTML]{E9FFE9}50.9}           & 4.4                     & 3.3                                   & 2.9           & \multicolumn{1}{c|}{2.2}                                   & 5.7           & \multicolumn{1}{c|}{\cellcolor[HTML]{E9FFE9}4.4}           & 3.0           & \multicolumn{1}{c|}{2.3}                                   & \textbf{11.7} & \multicolumn{1}{c|}{\cellcolor[HTML]{A6FFA6}\textbf{8.9}}  & \textbf{23.7} & \cellcolor[HTML]{A6FFA6}\textbf{18.1} \\
\multicolumn{1}{|l|}{}                               & SPR                                                        & 1.5           & \multicolumn{1}{c|}{\cellcolor[HTML]{FEFBCB}1.1}           & 3.5           & \multicolumn{1}{c|}{2.4}                                   & 4.0           & \multicolumn{1}{c|}{\cellcolor[HTML]{A6FFA6}\textbf{2.7}}  & 3.8           & \multicolumn{1}{c|}{\cellcolor[HTML]{A6FFA6}\textbf{2.6}}  & 37.1            & \multicolumn{1}{c|}{\cellcolor[HTML]{FEFBCB}25.2}           & 6.3                     & \cellcolor[HTML]{A6FFA6}\textbf{4.3}  & 1.7           & \multicolumn{1}{c|}{\cellcolor[HTML]{FEFBCB}1.1}           & 3.2           & \multicolumn{1}{c|}{\cellcolor[HTML]{FEFBCB}2.1}           & 2.5           & \multicolumn{1}{c|}{\cellcolor[HTML]{FEFBCB}1.7}           & 4.9           & \multicolumn{1}{c|}{\cellcolor[HTML]{FEFBCB}3.4}           & 19.1          & 13.0                                  \\
\multicolumn{1}{|l|}{}                               & Zen4                                                       & \textbf{4.9}  & \multicolumn{1}{c|}{\cellcolor[HTML]{A6FFA6}\textbf{2.9}}  & \textbf{6.1}  & \multicolumn{1}{c|}{\cellcolor[HTML]{A6FFA6}\textbf{3.6}}  & 4.4           & \multicolumn{1}{c|}{\cellcolor[HTML]{E9FFE9}2.6}           & 4.3           & \multicolumn{1}{c|}{\cellcolor[HTML]{A6FFA6}\textbf{2.6}}  & 56.6            & \multicolumn{1}{c|}{33.5}                                   & 4.0                     & 2.4                                   & \textbf{4.3}  & \multicolumn{1}{c|}{\cellcolor[HTML]{E9FFE9}2.5}           & 3.6           & \multicolumn{1}{c|}{\cellcolor[HTML]{FEFBCB}2.1}           & \textbf{6.5}  & \multicolumn{1}{c|}{\cellcolor[HTML]{A6FFA6}\textbf{3.9}}  & 9.0           & \multicolumn{1}{c|}{5.3}                                   & 21.6          & 12.8                                  \\
\multicolumn{1}{|l|}{\multirow{-6}{*}{\rotatebox[origin=c]{90}{\raisebox{-0.25\normalbaselineskip}[2pt][2pt]{\small arXiv/768}}}}    & SPR Z                                                      & 2.3           & \multicolumn{1}{c|}{\cellcolor[HTML]{FEFBCB}1.1}           & 3.9           & \multicolumn{1}{c|}{1.9}                                   & \textbf{4.8}  & \multicolumn{1}{c|}{\cellcolor[HTML]{E9FFE9}2.3}           & \textbf{4.5}  & \multicolumn{1}{c|}{\cellcolor[HTML]{E9FFE9}2.2}           & 40.3            & \multicolumn{1}{c|}{\cellcolor[HTML]{FFBFBE}19.5}           & \textbf{7.2}            & \cellcolor[HTML]{E9FFE9}3.5           & 2.0           & \multicolumn{1}{c|}{\cellcolor[HTML]{FEFBCB}1.0}           & 3.7           & \multicolumn{1}{c|}{\cellcolor[HTML]{FEFBCB}1.8}           & 2.7           & \multicolumn{1}{c|}{\cellcolor[HTML]{FFBFBE}1.3}           & 5.4           & \multicolumn{1}{c|}{\cellcolor[HTML]{FFBFBE}2.6}           & 21.3          & 10.3                                  \\ \hline
\multicolumn{1}{|l|}{}                               & Graviton3                                                  & 42.3          & \multicolumn{1}{c|}{\cellcolor[HTML]{E9FFE9}35.5}          & 13.8          & \multicolumn{1}{c|}{\cellcolor[HTML]{FFBFBE}11.6}          & 9.0           & \multicolumn{1}{c|}{\cellcolor[HTML]{FFBFBE}7.6}           & 10.3          & \multicolumn{1}{c|}{\cellcolor[HTML]{FFBFBE}8.6}           & 1032.0          & \multicolumn{1}{c|}{\cellcolor[HTML]{E9FFE9}867.3}          & 96.7                    & \cellcolor[HTML]{E9FFE9}81.2          & 27.4          & \multicolumn{1}{c|}{\cellcolor[HTML]{A6FFA6}\textbf{23.0}} & 35.9          & \multicolumn{1}{c|}{\cellcolor[HTML]{E9FFE9}30.2}          & 26.1          & \multicolumn{1}{c|}{\cellcolor[HTML]{E9FFE9}21.9}          & 47.0          & \multicolumn{1}{c|}{\cellcolor[HTML]{E9FFE9}39.5}          & 50.1          & \cellcolor[HTML]{E9FFE9}42.1          \\
\multicolumn{1}{|l|}{}                               & Zen3                                                       & 39.6          & \multicolumn{1}{c|}{31.4}                                  & 44.5          & \multicolumn{1}{c|}{35.3}                                  & 23.0          & \multicolumn{1}{c|}{\cellcolor[HTML]{FEFBCB}18.2}          & 26.3          & \multicolumn{1}{c|}{20.8}                                  & 1246.9          & \multicolumn{1}{c|}{\cellcolor[HTML]{A6FFA6}\textbf{989.4}} & 105.5                   & \cellcolor[HTML]{A6FFA6}\textbf{83.7} & 20.3          & \multicolumn{1}{c|}{16.1}                                  & 25.5          & \multicolumn{1}{c|}{20.2}                                  & 24.1          & \multicolumn{1}{c|}{\cellcolor[HTML]{E9FFE9}19.1}          & 38.4          & \multicolumn{1}{c|}{30.5}                                  & 54.5          & \cellcolor[HTML]{E9FFE9}43.2          \\
\multicolumn{1}{|l|}{}                               & Graviton4                                                  & 44.9          & \multicolumn{1}{c|}{\cellcolor[HTML]{E9FFE9}34.3}          & 17.0          & \multicolumn{1}{c|}{\cellcolor[HTML]{FFBFBE}13.0}          & 11.2          & \multicolumn{1}{c|}{\cellcolor[HTML]{FFBFBE}8.5}           & 12.5          & \multicolumn{1}{c|}{\cellcolor[HTML]{FFBFBE}9.5}           & 1081.1          & \multicolumn{1}{c|}{\cellcolor[HTML]{E9FFE9}826.2}          & 108.6                   & \cellcolor[HTML]{E9FFE9}83.0          & 28.2          & \multicolumn{1}{c|}{\cellcolor[HTML]{E9FFE9}21.5}          & \textbf{40.4} & \multicolumn{1}{c|}{\cellcolor[HTML]{A6FFA6}\textbf{30.9}} & 24.9          & \multicolumn{1}{c|}{\cellcolor[HTML]{E9FFE9}19.0}          & \textbf{55.6} & \multicolumn{1}{c|}{\cellcolor[HTML]{A6FFA6}\textbf{42.5}} & \textbf{59.2} & \cellcolor[HTML]{A6FFA6}\textbf{45.2} \\
\multicolumn{1}{|l|}{}                               & SPR                                                        & 22.8          & \multicolumn{1}{c|}{\cellcolor[HTML]{FEFBCB}15.5}          & 48.4          & \multicolumn{1}{c|}{32.9}                                  & 54.1          & \multicolumn{1}{c|}{\cellcolor[HTML]{A6FFA6}\textbf{36.8}} & 53.8          & \multicolumn{1}{c|}{\cellcolor[HTML]{A6FFA6}\textbf{36.6}} & 1121.1          & \multicolumn{1}{c|}{762.4}                                  & 107.3                   & \cellcolor[HTML]{E9FFE9}73.0          & 17.0          & \multicolumn{1}{c|}{11.6}                                  & 26.1          & \multicolumn{1}{c|}{17.8}                                  & 25.8          & \multicolumn{1}{c|}{17.5}                                  & 39.8          & \multicolumn{1}{c|}{27.1}                                  & 55.7          & \cellcolor[HTML]{E9FFE9}37.9          \\
\multicolumn{1}{|l|}{}                               & Zen4                                                       & \textbf{66.7} & \multicolumn{1}{c|}{\cellcolor[HTML]{A6FFA6}\textbf{39.5}} & \textbf{76.1} & \multicolumn{1}{c|}{\cellcolor[HTML]{A6FFA6}\textbf{45.0}} & 55.4          & \multicolumn{1}{c|}{\cellcolor[HTML]{E9FFE9}32.8}          & 51.8          & \multicolumn{1}{c|}{\cellcolor[HTML]{E9FFE9}30.6}          & 1344.1          & \multicolumn{1}{c|}{\cellcolor[HTML]{E9FFE9}794.9}          & 108.8                   & 64.4                                  & \textbf{34.9} & \multicolumn{1}{c|}{\cellcolor[HTML]{E9FFE9}20.6}          & 25.8          & \multicolumn{1}{c|}{\cellcolor[HTML]{FEFBCB}15.3}          & \textbf{38.4} & \multicolumn{1}{c|}{\cellcolor[HTML]{A6FFA6}\textbf{22.7}} & 42.3          & \multicolumn{1}{c|}{25.0}                                  & 52.0          & 30.7                                  \\
\multicolumn{1}{|l|}{\multirow{-6}{*}{\rotatebox[origin=c]{90}{\raisebox{-0.25\normalbaselineskip}[2pt][2pt]{\small SIFT/128}}}}     & SPR Z                                                      & 26.9          & \multicolumn{1}{c|}{\cellcolor[HTML]{FFBFBE}13.0}          & 54.6          & \multicolumn{1}{c|}{26.4}                                  & \textbf{59.8} & \multicolumn{1}{c|}{29.0}                                  & \textbf{56.8} & \multicolumn{1}{c|}{27.5}                                  & \textbf{1396.6} & \multicolumn{1}{c|}{675.5}                                  & \textbf{112.8}          & 54.6                                  & 19.0          & \multicolumn{1}{c|}{\cellcolor[HTML]{FEFBCB}9.2}           & 28.9          & \multicolumn{1}{c|}{14.0}                                  & 29.0          & \multicolumn{1}{c|}{14.0}                                  & 44.0          & \multicolumn{1}{c|}{21.3}                                  & 58.7          & 28.4                                  \\ \hline
\end{tabular}
}
%\vspace*{-3mm}
\end{table*}

\subsection{HNSW}\label{sec:eval-hnsw}
Figure~\ref{fig:hnsw} shows the performance of the different microarchitectures on an HNSW index without quantization. In this setting, Intel $Z$ offers the highest QPS. However, it never gives the highest QP\$. For QP\$, Zen3 and Graviton3 perform the best in high-dimensional vectors. Contrary to IVF indexes, Zen4 performs the worst. Another observation is that, like in IVF indexes, the targeted recall does not affect the relative performance of microarchitectures. 

Table~\ref{tab:hnsw} shows each microarchitecture's performance with different quantization levels at the highest possible recall. Intel $Z$ gives the highest QPS in nearly all settings. However, things change when looking at QP\$. In {\tt \small float16} vectors, the Intels and Graviton3 shine, and the Zens struggle due to the lack of {\tt \small float16} SIMD. On {\tt \small bfloat}, the Gravitons are the winners for high-dimensional vectors, with Zen3 following closely, despite the latter not having direct support for {\tt \small bfloat}. Finally, in quantized vectors, the Gravitons give the highest QP\$ at higher dimensionalities, and the Intels on vectors of lower dimensionality. It is important to note that in HNSW, the gap between architectures is less evident (fewer red and yellow cells). 

\subsection{Full Scans}\label{sec:eval-full}
Table~\ref{tab:full} shows the performance of each microarchitecture when doing full scans. For {\tt \small float32} vectors, Zen4 and Graviton3 give the best QP\$, while the Intels fall short. Contrary to HNSW, the Gravitons always take the lead on {\tt \small float16} instead of Intel, and Zen4 takes the upper hand on {\tt \small bfloat}. In 8-bit vectors, USearch and FAISS differ: the Gravitons take the lead in USearch thanks to the symmetric kernels, whereas in FAISS, Zen4 and Intel perform well across SQ settings, with Gravitons heavily underperforming again (lack of SIMD decoding). Finally, on 1-bit and PQ vectors, the Gravitons and Intel are the best performers, respectively.

\begin{figure*}[t!]
\centering
\includegraphics[width=1.0\linewidth]{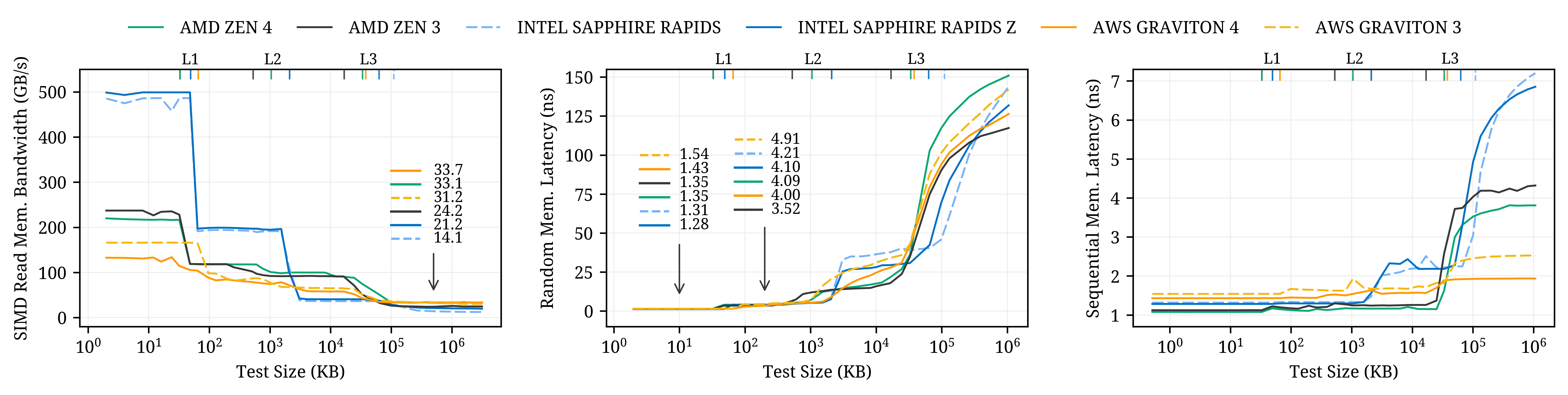}
\vspace*{-8mm}
\caption{SIMD read bandwidth and access latency (random and sequential) of cache and DRAM in cloud CPUs. %Non-quantized IVF and full scans are benefited by higher bandwidth at L3/DRAM and lower latencies on sequential access. Quantized indexes and HNSW are benefited by higher bandwidth at lower memory levels and lower latencies on random access. 
}
\label{fig:memory}
\vspace*{-3mm}
\end{figure*}

\subsection{Why the differences across microarchs?}\label{sec:why}

While having symmetric kernels with specialized SIMD \textit{does} make a difference in performance under certain settings, it is not a rule of thumb for which microarchitecture will perform better. For instance, SPR does not win in {\small \tt float16}+full scan. The performance differences arise due to two factors: the data-access patterns of the search algorithm (since vector search is mostly data-access bound~\cite{pdx, sancaaccess, graphmembounded}) and the efficiency of the distance kernel. 

%\vspace*{3mm}
\noindent{\bf Data-access patterns of indexes. } On the one hand, IVF indexes (and full scans) sequentially access large chunks of vectors that likely exceed the size of the L2 cache or L3 in full scans. As a result, microarchitectures with higher latencies and lower read bandwidths at L3/DRAM are at a disadvantage. Our memory latency benchmarks, presented in Figure~\ref{fig:memory}, show that when data is bigger than L3, the latency of sequential access (plot on the right) on SPR is almost 2x higher than the Zens and almost 4x higher than the Gravitons. Furthermore, the memory bandwidth (plot on the left) of SPR at L3 and DRAM is the lowest. These data-access-related capabilities are reflected in our vector search experiments, in which the performance gap between SPR and Zen4 closes in IVF and full scans as vectors are quantized at smaller bit-widths (see in Table ~\ref{tab:ivf} how SPR becomes green from {\small \tt float32} to SQ4). 

% These observations align with benchmarks reported in other studies~\cite{microfog, intelmemoryaccess, chipsgraviton, chipssapphire}. 

On the other hand, searches on an HNSW index access fewer vectors, which results in less pressure on the caches. Moreover, the vectors in the upper layer of the HNSW index are cached efficiently as the entry point of the index is always the same (see left of Figure~\ref{fig:ivf-hnsw}). Hereby, microarchitectures like SPR take advantage, especially with quantized vectors, thanks to having bigger L2/L3 (see Table~\ref{tab:machine_details}), higher read bandwidth at L1/L2, and a higher SIMD {\tt \small LOAD} throughput. These observations align with our HNSW experiments, where both SPRs have the highest QPS in all scenarios. PQ also benefits from architectures like SPR, as the codebooks can be cached and accessed more efficiently.

\vspace*{3mm}
\noindent{\bf Efficiency of the distance kernel. } The raw performance of distance kernels depends on the execution throughput and latency of the SIMD instructions used. We define \textit{execution throughput} as the maximum number of bits an instruction can process per CPU cycle across the microarchitecture~\cite{armv1}. Both SPR and Zen4 have a floating-point (FP) execution throughput of 1024 bits per CPU cycle~\cite{microfog}. On the other hand, the Gravitons have an FP execution throughput of 512 bits, which is half of Zen4 and SPR. However, the 1024 bits of Zen4 are arranged as four specialized execution ports of 256 bits each, two for {\small \tt FP-FMA} (fused multiply-add) and two for {\small \tt FP-ADD}. Effectively achieving a maximum throughput of 512 bits per cycle for {\small \tt FP-FMA}. On the contrary, all four ports (512 bits) in Gravitons can handle both FP operations. Thus, depending on the distance kernel, the Gravitons can be on par with Zen4 in FP instructions throughput. However, as data must be first \textit{loaded} into the registers, the throughput of {\small \tt LOAD} instructions also plays a key role in the efficiency of the kernels. In fact, SPR's main advantage on SIMD is that it can serve 2x512 bits loads per cycle from L1, while Zen4 and Graviton3 can only load 2x256 bits~\cite{microfog,armv1} and Graviton4 3x128 bits~\cite{armv2}. This effect is visible in the left plot of Figure~\ref{fig:memory}, with SPR having twice as much bandwidth than the Zens when data fits in L1. However, this advantage rapidly degrades as soon as data spills to L3. Finally, regarding latencies, all architectures are close by, with latencies of 3-4 cycles for FP  instructions~\cite{microfog, uops, chipsgraviton, chipssapphire, armv1, armv2}. However, at smaller register widths, the instructions must be called more times to process the same amount of values. The latter gives an advantage to wider register widths due to the instructions call latency.  

\vspace*{3mm}
\noindent{\bf Data size. } To further investigate the effect of data size, we ran full scans on random collections of {\small \tt float32} vectors of different sizes and dimensionalities. We found that when the collection fits in the L2 cache, SPR achieves 10\% more performance than Zen4, with Zen3 and the Gravitons underperforming. However, as soon as data spills to L3, the performance of SPR degrades, and Zen4 takes the lead. SPR performance further degrades when data spills to DRAM, delivering 30\% less performance than Zen3 and Graviton4 and 2x less performance than Zen4.

\vspace*{3mm}
\noindent{\bf Graviton4 vs Graviton3.} In many of our experiments, Graviton3 performed better than Graviton4, both in QP\$ and QPS (see {\small \tt float32} in Table \ref{tab:ivf}). The main reason behind this is that Graviton3 has double the size of SVE registers (256 bits) compared to Graviton4 (128 bits). While both architectures have the same FP execution throughput, the total latency cost to process the same amount of data is higher in Graviton4 due to the smaller register. The latter becomes more critical in distance kernels used in vector search, where a dependency chain exists as distances are accumulated on the same SIMD lanes. Furthermore, Graviton3 has a higher execution throughput of {\tt \small LOAD}, capable of loading 2x256 bits in one CPU cycle~\cite{armv1}, while Graviton4 only 3x128 bits~\cite{armv2}.  Further experiments on full scans on our random collections of {\tt \small float32} vectors reveal that SVE is 37\% faster than NEON on Graviton3 (due to doubling the register size). In contrast, in Graviton4, NEON and SVE perform the same since the register size, execution throughput, and latencies are the same. When comparing NEON to NEON, Graviton4 is faster than Graviton3 by 10\%. However, when switching to SVE, the tables turn, and Graviton3 is 31\% faster than Graviton4. These findings differ from most benchmarks, in which Graviton4 always delivers more performance. However, we believe these observations have been under the radar since most benchmarks use NEON, as SVE has not yet been widely adopted.

% All in all, the speed of distance kernels boils down to the CPU architectural design. However, in the bigger picture of VSS, it is an interplay between the memory-access patterns of the search algorithm, the size of the vectors (given by the quantization technique and dimensionality), and the speed of the distance kernel. 

\section{Bang for the buck: Takeways}\label{sec:summary}
In Table~\ref{tab:summary}, we rank every microarchitecture on five tiers based on their QPS and QP\$ on the OpenAI/1536 dataset: (\colorbox[HTML]{A6FFA6}{++}) for the best one, (\colorbox[HTML]{E9FFE9}{+}) for the ones with a performance at most 20\% away from the best, (\colorbox[HTML]{fefbcb}{-}) when they provide 2x less, and (\colorbox[HTML]{ffbfbe}{-{}-}) when they provide 3x less QP\$ than the best option, and (\colorbox[HTML]{f5f5f5}{$\cdot$}) for options within the middle ground. Finally, we present an aggregated score by giving 1 point for every (+) and -1 point for every (-). 

Graviton3 gives the best ``bang for the buck,'' even over its successor, Graviton4. Graviton3 is only pushed back in our scoring system due to the lack of symmetric kernels in FAISS. Note that Graviton3 excels in the following areas: a variety of SIMD capabilities, high read throughput, and low sequential memory latency at L3/DRAM. More importantly, it is cheap. 
Zen4 is still a solid option for vector search on IVF indexes and full scans, especially in {\tt \small float32} and {\tt \small bfloat} vectors. Zen3 has few negative points thanks to its low price and low latencies for L2/L3/DRAM access. However, it does not excel in any setting. Finally, the SPRs have the lowest score in QP\$ and do not excel in any setting. Despite SPR Z having the best QPS score for HNSW, its price brings down its QP\$ score. 

% \textcolor{blue}{}

% All in all, the speed of vector search is given by an interplay between the memory-access patterns of the search algorithm, the size of the vectors (given by the quantization technique and dimensionality), and the speed of the distance kernel. 

%\vspace*{3mm}
%\noindent{\bf Key takeaway:} For IVF indexes or no index, use Zen4 (high L3/DRAM bandwidth and low sequential memory latency). For quantized vectors or HNSW indexes, use Graviton 3 for the best ``bang for the buck'', and, if money is not a concern, use Intel Z to get the best performance (big caches and high L1/L2 bandwidths). If the use case is not clear, use Graviton 3.

\section{Discussion}\label{sec:discussion}

The need for Vector Databases is widely disputed among the community~\cite{lucene}, with some foreseeing them merging with existing database systems. While the future of VecDBs as standalone systems remains uncertain, vector workloads are here to stay. We believe our insights are important as the decoupling of storage and compute that the cloud provides makes it possible to switch microarchitectures easily depending on the search algorithm to be used. Furthermore, providing the right microarchitecture can incur huge savings, especially in serverless vector search services~\cite{vexless, squash}, where every millisecond counts toward billing. 

In this study, we have focused on the three most common use cases in vector search: HNSW, IVF, and full scans, at different quantization levels. However, there exists a wider variety of indexes (e.g., hybrids), quantization techniques (e.g., residual quantizers~\cite{faisspaper}), and distance metrics (e.g., cosine similarity, inner product). Therefore, we encourage users to perform data-driven benchmarks to uncover the best microarchitecture for their use case.

It is important to acknowledge that the microarchitectures presented in this study will be outdated in the future. For instance, although not yet available in AWS, Intel has already released Emerald Rapids, the successor to Sapphire Rapids. Similarly, AMD has already released Zen5, the successor to Zen4. Nevertheless, the insights presented here should motivate researchers in the vector search community to not only strive for lower theoretical complexities but also for better data-access patterns and storage designs of newly developed algorithms. The latter is critical for the performance of vector search on a large scale, where it is heavily data-access bound~\cite{scannwhitepaper, sancaaccess, pdx, turbolvq, rabitqext}. Finally, we encourage researchers \textit{always} to use SIMD-optimized implementations, as newly developed algorithms that may shine in their scalar implementation can fail to be on par with SIMD-optimized approaches~\cite{pdx, tribase}.

\vfill

\section{Conclusions and Future Work}\label{sec:conclusion}
In this paper, we have shed light on how to get the best ``bang for the buck'' when using vector search solutions in the cloud. We showed that the performance of vector search does not solely depend on the SIMD capabilities of the underlying CPU, as the data-access patterns of the search algorithm and the size of the vectors also play an important role. We have shown that these differences in CPUs and search algorithms are enough to get $\approx$3x more queries per dollar (QP\$) if the microarchitecture is carefully chosen. Overall, in AWS, Graviton3 (r7g) and Zen4 (r7a) CPUs give you a good ``bang for the buck'' in the majority of vector search scenarios.

In future work, different types of vector-centric workloads can be benchmarked (e.g., interleaving updates and queries, batch queries with multithreading, index construction). Also, a similar study with different GPU microarchitectures could bring valuable insights. Furthermore, a study comparing the QP\$ yielded by different VecDBs would be a next step to understand further the tradeoff between storage designs and distance kernels implementations. Ultimately, the current landscape of vector research needs more established benchmarks, such as TPC-H or TPC-DS in conventional databases.

\iffalse
\begin{acks}
...
\end{acks}
\fi

\bibliographystyle{ACM-Reference-Format}
\bibliography{_main}


\begin{thebibliography}{49}

%%% the Web address.

\ifx \showCODEN    \undefined \def \showCODEN     #1{\unskip}     \fi
\ifx \showDOI      \undefined \def \showDOI       #1{#1}\fi
\ifx \showISBNx    \undefined \def \showISBNx     #1{\unskip}     \fi
\ifx \showISBNxiii \undefined \def \showISBNxiii  #1{\unskip}     \fi
\ifx \showISSN     \undefined \def \showISSN      #1{\unskip}     \fi
\ifx \showLCCN     \undefined \def \showLCCN      #1{\unskip}     \fi
\ifx \shownote     \undefined \def \shownote      #1{#1}          \fi
\ifx \showarticletitle \undefined \def \showarticletitle #1{#1}   \fi
\ifx \showURL      \undefined \def \showURL       {\relax}        \fi
% invisible to TeX
\providecommand\bibfield[2]{#2}
\providecommand\bibinfo[2]{#2}
\providecommand\natexlab[1]{#1}
\providecommand\showeprint[2][]{arXiv:#2}

\bibitem[Abel and Reineke(2019)]%
        {uops}
\bibfield{author}{\bibinfo{person}{Andreas Abel} {and} \bibinfo{person}{Jan Reineke}.} \bibinfo{year}{2019}\natexlab{}.
\newblock \showarticletitle{uops. info: Characterizing latency, throughput, and port usage of instructions on intel microarchitectures}. In \bibinfo{booktitle}{\emph{Proceedings of the Twenty-Fourth International Conference on Architectural Support for Programming Languages and Operating Systems}}. \bibinfo{pages}{673--686}.
\newblock


\bibitem[Afroozeh and Boncz(2023)]%
        {fastlanes}
\bibfield{author}{\bibinfo{person}{Azim Afroozeh} {and} \bibinfo{person}{Peter Boncz}.} \bibinfo{year}{2023}\natexlab{}.
\newblock \showarticletitle{The FastLanes Compression Layout: Decoding \&gt; 100 Billion Integers per Second with Scalar Code}.
\newblock \bibinfo{journal}{\emph{Proc. VLDB Endow.}} \bibinfo{volume}{16}, \bibinfo{number}{9} (\bibinfo{date}{jul} \bibinfo{year}{2023}), \bibinfo{pages}{2132–2144}.
\newblock
\showISSN{2150-8097}
\urldef\tempurl%
\url{https://doi.org/10.14778/3598581.3598587}
\showDOI{\tempurl}


\bibitem[Aguerrebere et~al\mbox{.}(2023)]%
        {lvq}
\bibfield{author}{\bibinfo{person}{Cecilia Aguerrebere}, \bibinfo{person}{Ishwar Bhati}, \bibinfo{person}{Mark Hildebrand}, \bibinfo{person}{Mariano Tepper}, {and} \bibinfo{person}{Ted Willke}.} \bibinfo{year}{2023}\natexlab{}.
\newblock \showarticletitle{Similarity search in the blink of an eye with compressed indices}.
\newblock \bibinfo{journal}{\emph{arXiv preprint arXiv:2304.04759}} (\bibinfo{year}{2023}).
\newblock


\bibitem[Aguerrebere et~al\mbox{.}(2024)]%
        {turbolvq}
\bibfield{author}{\bibinfo{person}{Cecilia Aguerrebere}, \bibinfo{person}{Mark Hildebrand}, \bibinfo{person}{Ishwar~Singh Bhati}, \bibinfo{person}{Theodore Willke}, {and} \bibinfo{person}{Mariano Tepper}.} \bibinfo{year}{2024}\natexlab{}.
\newblock \showarticletitle{Locally-Adaptive Quantization for Streaming Vector Search}.
\newblock \bibinfo{journal}{\emph{arXiv preprint arXiv:2402.02044}} (\bibinfo{year}{2024}).
\newblock


\bibitem[{Arm Limited}(2022a)]%
        {armv1}
\bibfield{author}{\bibinfo{person}{{Arm Limited}}.} \bibinfo{year}{2022}\natexlab{a}.
\newblock \bibinfo{booktitle}{\emph{Arm Neoverse V1 Software Optimization Guide}}.
\newblock
\newblock
\shownote{Version 6.0}.


\bibitem[{Arm Limited}(2022b)]%
        {armv2}
\bibfield{author}{\bibinfo{person}{{Arm Limited}}.} \bibinfo{year}{2022}\natexlab{b}.
\newblock \bibinfo{booktitle}{\emph{Arm Neoverse V2 Software Optimization Guide}}.
\newblock
\newblock
\shownote{Version 3.0}.


\bibitem[Aum{\"u}ller et~al\mbox{.}(2020)]%
        {annbench}
\bibfield{author}{\bibinfo{person}{Martin Aum{\"u}ller}, \bibinfo{person}{Erik Bernhardsson}, {and} \bibinfo{person}{Alexander Faithfull}.} \bibinfo{year}{2020}\natexlab{}.
\newblock \showarticletitle{ANN-Benchmarks: A benchmarking tool for approximate nearest neighbor algorithms}.
\newblock \bibinfo{journal}{\emph{Information Systems}}  \bibinfo{volume}{87} (\bibinfo{year}{2020}), \bibinfo{pages}{101374}.
\newblock


\bibitem[Chen et~al\mbox{.}(2021)]%
        {spann}
\bibfield{author}{\bibinfo{person}{Qi Chen}, \bibinfo{person}{Bing Zhao}, \bibinfo{person}{Haidong Wang}, \bibinfo{person}{Mingqin Li}, \bibinfo{person}{Chuanjie Liu}, \bibinfo{person}{Zengzhong Li}, \bibinfo{person}{Mao Yang}, {and} \bibinfo{person}{Jingdong Wang}.} \bibinfo{year}{2021}\natexlab{}.
\newblock \showarticletitle{Spann: Highly-efficient billion-scale approximate nearest neighborhood search}.
\newblock \bibinfo{journal}{\emph{Advances in Neural Information Processing Systems}}  \bibinfo{volume}{34} (\bibinfo{year}{2021}), \bibinfo{pages}{5199--5212}.
\newblock


\bibitem[Chroma(2024)]%
        {chromaperf}
\bibfield{author}{\bibinfo{person}{Chroma}.} \bibinfo{year}{2024}\natexlab{}.
\newblock \bibinfo{title}{Single-Node Chroma: Performance and Limitations}.
\newblock \bibinfo{howpublished}{\url{https://docs.trychroma.com/production/administration/performance}}.
\newblock


\bibitem[David~Myriel(2024)]%
        {intelqdrant}
\bibfield{author}{\bibinfo{person}{Kumar~Shivendu David~Myriel}.} \bibinfo{year}{2024}\natexlab{}.
\newblock \bibinfo{title}{Intel’s New CPU Powers Faster Vector Search}.
\newblock \bibinfo{howpublished}{\url{https://qdrant.tech/blog/qdrant-cpu-intel-benchmark/}}.
\newblock


\bibitem[Douze et~al\mbox{.}(2024)]%
        {faisspaper}
\bibfield{author}{\bibinfo{person}{Matthijs Douze}, \bibinfo{person}{Alexandr Guzhva}, \bibinfo{person}{Chengqi Deng}, \bibinfo{person}{Jeff Johnson}, \bibinfo{person}{Gergely Szilvasy}, \bibinfo{person}{Pierre-Emmanuel Mazar{\'e}}, \bibinfo{person}{Maria Lomeli}, \bibinfo{person}{Lucas Hosseini}, {and} \bibinfo{person}{Herv{\'e} J{\'e}gou}.} \bibinfo{year}{2024}\natexlab{}.
\newblock \showarticletitle{The faiss library}.
\newblock \bibinfo{journal}{\emph{arXiv preprint arXiv:2401.08281}} (\bibinfo{year}{2024}).
\newblock


\bibitem[Fog(2016)]%
        {microfog}
\bibfield{author}{\bibinfo{person}{Agner Fog}.} \bibinfo{year}{2016}\natexlab{}.
\newblock \showarticletitle{The microarchitecture of Intel, AMD and VIA CPUs: An optimization guide for assembly programmers and compiler makers}.
\newblock \bibinfo{journal}{\emph{Software optimization resources}} (\bibinfo{year}{2016}).
\newblock


\bibitem[Gao et~al\mbox{.}(2024)]%
        {rabitqext}
\bibfield{author}{\bibinfo{person}{Jianyang Gao}, \bibinfo{person}{Yutong Gou}, \bibinfo{person}{Yuexuan Xu}, \bibinfo{person}{Yongyi Yang}, \bibinfo{person}{Cheng Long}, {and} \bibinfo{person}{Raymond Chi-Wing Wong}.} \bibinfo{year}{2024}\natexlab{}.
\newblock \showarticletitle{Practical and Asymptotically Optimal Quantization of High-Dimensional Vectors in Euclidean Space for Approximate Nearest Neighbor Search}.
\newblock \bibinfo{journal}{\emph{arXiv preprint arXiv:2409.09913}} (\bibinfo{year}{2024}).
\newblock


\bibitem[Gao and Long(2023)]%
        {adsampling}
\bibfield{author}{\bibinfo{person}{Jianyang Gao} {and} \bibinfo{person}{Cheng Long}.} \bibinfo{year}{2023}\natexlab{}.
\newblock \showarticletitle{High-dimensional approximate nearest neighbor search: with reliable and efficient distance comparison operations}.
\newblock \bibinfo{journal}{\emph{Proceedings of the ACM on Management of Data}} \bibinfo{volume}{1}, \bibinfo{number}{2} (\bibinfo{year}{2023}), \bibinfo{pages}{1--27}.
\newblock


\bibitem[Guo et~al\mbox{.}(2020)]%
        {anisotropic}
\bibfield{author}{\bibinfo{person}{Ruiqi Guo}, \bibinfo{person}{Philip Sun}, \bibinfo{person}{Erik Lindgren}, \bibinfo{person}{Quan Geng}, \bibinfo{person}{David Simcha}, \bibinfo{person}{Felix Chern}, {and} \bibinfo{person}{Sanjiv Kumar}.} \bibinfo{year}{2020}\natexlab{}.
\newblock \showarticletitle{Accelerating large-scale inference with anisotropic vector quantization}. In \bibinfo{booktitle}{\emph{International Conference on Machine Learning}}. PMLR, \bibinfo{pages}{3887--3896}.
\newblock


\bibitem[Iwasaki and Miyazaki(2018)]%
        {ngt}
\bibfield{author}{\bibinfo{person}{Masajiro Iwasaki} {and} \bibinfo{person}{Daisuke Miyazaki}.} \bibinfo{year}{2018}\natexlab{}.
\newblock \showarticletitle{Optimization of indexing based on k-nearest neighbor graph for proximity search in high-dimensional data}.
\newblock \bibinfo{journal}{\emph{arXiv preprint arXiv:1810.07355}} (\bibinfo{year}{2018}).
\newblock


\bibitem[Jafari et~al\mbox{.}(2021)]%
        {surveylsh}
\bibfield{author}{\bibinfo{person}{Omid Jafari}, \bibinfo{person}{Preeti Maurya}, \bibinfo{person}{Parth Nagarkar}, \bibinfo{person}{Khandker~Mushfiqul Islam}, {and} \bibinfo{person}{Chidambaram Crushev}.} \bibinfo{year}{2021}\natexlab{}.
\newblock \showarticletitle{A survey on locality sensitive hashing algorithms and their applications}.
\newblock \bibinfo{journal}{\emph{arXiv preprint arXiv:2102.08942}} (\bibinfo{year}{2021}).
\newblock


\bibitem[Jegou et~al\mbox{.}(2010)]%
        {pqivf}
\bibfield{author}{\bibinfo{person}{Herve Jegou}, \bibinfo{person}{Matthijs Douze}, {and} \bibinfo{person}{Cordelia Schmid}.} \bibinfo{year}{2010}\natexlab{}.
\newblock \showarticletitle{Product quantization for nearest neighbor search}.
\newblock \bibinfo{journal}{\emph{IEEE transactions on pattern analysis and machine intelligence}} \bibinfo{volume}{33}, \bibinfo{number}{1} (\bibinfo{year}{2010}), \bibinfo{pages}{117--128}.
\newblock


\bibitem[Jing et~al\mbox{.}(2024)]%
        {llmmeetvdbs}
\bibfield{author}{\bibinfo{person}{Zhi Jing}, \bibinfo{person}{Yongye Su}, \bibinfo{person}{Yikun Han}, \bibinfo{person}{Bo Yuan}, \bibinfo{person}{Haiyun Xu}, \bibinfo{person}{Chunjiang Liu}, \bibinfo{person}{Kehai Chen}, {and} \bibinfo{person}{Min Zhang}.} \bibinfo{year}{2024}\natexlab{}.
\newblock \showarticletitle{When large language models meet vector databases: A survey}.
\newblock \bibinfo{journal}{\emph{arXiv preprint arXiv:2402.01763}} (\bibinfo{year}{2024}).
\newblock


\bibitem[Ko et~al\mbox{.}(2021)]%
        {lowquant}
\bibfield{author}{\bibinfo{person}{Anthony Ko}, \bibinfo{person}{Iman Keivanloo}, \bibinfo{person}{Vihan Lakshman}, {and} \bibinfo{person}{Eric Schkufza}.} \bibinfo{year}{2021}\natexlab{}.
\newblock \showarticletitle{Low-precision quantization for efficient nearest neighbor search}.
\newblock \bibinfo{journal}{\emph{arXiv preprint arXiv:2110.08919}} (\bibinfo{year}{2021}).
\newblock


\bibitem[Kuffo et~al\mbox{.}(2025)]%
        {pdx}
\bibfield{author}{\bibinfo{person}{Leonardo Kuffo}, \bibinfo{person}{Elena Krippner}, {and} \bibinfo{person}{Peter Boncz}.} \bibinfo{year}{2025}\natexlab{}.
\newblock \showarticletitle{{PDX:} A Data Layout for Vector Similarity Search}.
\newblock \bibinfo{journal}{\emph{Proc. {ACM} Manag. Data}} (\bibinfo{year}{2025}).
\newblock


\bibitem[Lam(2023)]%
        {chipssapphire}
\bibfield{author}{\bibinfo{person}{Chester Lam}.} \bibinfo{year}{2023}\natexlab{}.
\newblock \bibinfo{title}{Sapphire Rapids: Golden Cove Hits Servers}.
\newblock \bibinfo{howpublished}{\url{https://chipsandcheese.com/p/a-peek-at-sapphire-rapids}}.
\newblock


\bibitem[Lam(2024)]%
        {chipsgraviton}
\bibfield{author}{\bibinfo{person}{Chester Lam}.} \bibinfo{year}{2024}\natexlab{}.
\newblock \bibinfo{title}{Arm’s Neoverse V2, in AWS’s Graviton 4}.
\newblock \bibinfo{howpublished}{\url{https://chipsandcheese.com/p/arms-neoverse-v2-in-awss-graviton-4}}.
\newblock


\bibitem[Lemire(2018)]%
        {lemireavx512bad}
\bibfield{author}{\bibinfo{person}{Daniel Lemire}.} \bibinfo{year}{2018}\natexlab{}.
\newblock \bibinfo{title}{AVX-512: when and how to use these new instructions}.
\newblock \bibinfo{howpublished}{\url{https://lemire.me/blog/2018/09/07/avx-512-when-and-how-to-use-these-new-instructions/}}.
\newblock


\bibitem[Malkov and Yashunin(2018)]%
        {hnsw}
\bibfield{author}{\bibinfo{person}{Yu~A Malkov} {and} \bibinfo{person}{Dmitry~A Yashunin}.} \bibinfo{year}{2018}\natexlab{}.
\newblock \showarticletitle{Efficient and robust approximate nearest neighbor search using hierarchical navigable small world graphs}.
\newblock \bibinfo{journal}{\emph{IEEE transactions on pattern analysis and machine intelligence}} \bibinfo{volume}{42}, \bibinfo{number}{4} (\bibinfo{year}{2018}), \bibinfo{pages}{824--836}.
\newblock


\bibitem[Milvus(2020)]%
        {milvusavx}
\bibfield{author}{\bibinfo{person}{zilliz Milvus}.} \bibinfo{year}{2020}\natexlab{}.
\newblock \bibinfo{title}{Milvus performance on AVX-512 vs. AVX2}.
\newblock \bibinfo{howpublished}{\url{https://milvus.io/blog/2020-11-10-milvus-performance-AVX-512-vs-AVX2.md}}.
\newblock


\bibitem[Munoz et~al\mbox{.}(2019)]%
        {hcnng}
\bibfield{author}{\bibinfo{person}{Javier~Vargas Munoz}, \bibinfo{person}{Marcos~A Gon{\c{c}}alves}, \bibinfo{person}{Zanoni Dias}, {and} \bibinfo{person}{Ricardo da~S Torres}.} \bibinfo{year}{2019}\natexlab{}.
\newblock \showarticletitle{Hierarchical clustering-based graphs for large scale approximate nearest neighbor search}.
\newblock \bibinfo{journal}{\emph{Pattern Recognition}}  \bibinfo{volume}{96} (\bibinfo{year}{2019}), \bibinfo{pages}{106970}.
\newblock


\bibitem[Oakley and Ferhatosmanoglu(2025)]%
        {squash}
\bibfield{author}{\bibinfo{person}{Joe Oakley} {and} \bibinfo{person}{Hakan Ferhatosmanoglu}.} \bibinfo{year}{2025}\natexlab{}.
\newblock \showarticletitle{SQUASH: Serverless and Distributed Quantization-based Attributed Vector Similarity Search}.
\newblock \bibinfo{journal}{\emph{arXiv preprint arXiv:2502.01528}} (\bibinfo{year}{2025}).
\newblock


\bibitem[Pan et~al\mbox{.}(2023)]%
        {surveysystems}
\bibfield{author}{\bibinfo{person}{James~Jie Pan}, \bibinfo{person}{Jianguo Wang}, {and} \bibinfo{person}{Guoliang Li}.} \bibinfo{year}{2023}\natexlab{}.
\newblock \showarticletitle{Survey of vector database management systems}.
\newblock \bibinfo{journal}{\emph{arXiv preprint arXiv:2310.14021}} (\bibinfo{year}{2023}).
\newblock


\bibitem[Pan et~al\mbox{.}(2024)]%
        {vectordbs}
\bibfield{author}{\bibinfo{person}{James~Jie Pan}, \bibinfo{person}{Jianguo Wang}, {and} \bibinfo{person}{Guoliang Li}.} \bibinfo{year}{2024}\natexlab{}.
\newblock \showarticletitle{Vector Database Management Techniques and Systems}. In \bibinfo{booktitle}{\emph{Companion of the 2024 International Conference on Management of Data}}. \bibinfo{pages}{597--604}.
\newblock


\bibitem[Papakonstantinou et~al\mbox{.}(2024)]%
        {scannwhitepaper}
\bibfield{author}{\bibinfo{person}{Yannis Papakonstantinou}, \bibinfo{person}{Alan Li}, \bibinfo{person}{Ruiqi Guo}, \bibinfo{person}{Sanjiv Kumar}, {and} \bibinfo{person}{Phil Sun}.} \bibinfo{year}{2024}\natexlab{}.
\newblock \bibinfo{booktitle}{\emph{ScaNN for AlloyDB}}.
\newblock \bibinfo{type}{{T}echnical {R}eport}. \bibinfo{institution}{Google Cloud}.
\newblock
\urldef\tempurl%
\url{https://services.google.com/fh/files/misc/scann_for_alloydb_whitepaper.pdf}
\showURL{%
\tempurl}
\newblock
\shownote{Whitepaper}.


\bibitem[Patel et~al\mbox{.}(2024)]%
        {patel2024acorn}
\bibfield{author}{\bibinfo{person}{Liana Patel}, \bibinfo{person}{Peter Kraft}, \bibinfo{person}{Carlos Guestrin}, {and} \bibinfo{person}{Matei Zaharia}.} \bibinfo{year}{2024}\natexlab{}.
\newblock \showarticletitle{Acorn: Performant and predicate-agnostic search over vector embeddings and structured data}.
\newblock \bibinfo{journal}{\emph{Proceedings of the ACM on Management of Data}} \bibinfo{volume}{2}, \bibinfo{number}{3} (\bibinfo{year}{2024}), \bibinfo{pages}{1--27}.
\newblock


\bibitem[Research(2024)]%
        {faisscode}
\bibfield{author}{\bibinfo{person}{Meta Research}.} \bibinfo{year}{2024}\natexlab{}.
\newblock \bibinfo{booktitle}{\emph{{Faiss: A library for efficient similarity search and clustering of dense vectors.}}}
\newblock
\urldef\tempurl%
\url{https://github.com/facebookresearch/faiss}
\showURL{%
\tempurl}


\bibitem[Sanca and Ailamaki(2024)]%
        {sancaaccess}
\bibfield{author}{\bibinfo{person}{Viktor Sanca} {and} \bibinfo{person}{Anastasia Ailamaki}.} \bibinfo{year}{2024}\natexlab{}.
\newblock \showarticletitle{Efficient Data Access Paths for Mixed Vector-Relational Search}. In \bibinfo{booktitle}{\emph{Proceedings of the 20th International Workshop on Data Management on New Hardware}}. \bibinfo{pages}{1--9}.
\newblock


\bibitem[Spotify(2017)]%
        {annoy}
\bibfield{author}{\bibinfo{person}{Spotify}.} \bibinfo{year}{2017}\natexlab{}.
\newblock \bibinfo{booktitle}{\emph{{ANNOY by Spotify}}}.
\newblock
\urldef\tempurl%
\url{https://github.com/spotify/annoy}
\showURL{%
\tempurl}


\bibitem[Su et~al\mbox{.}(2024)]%
        {vexless}
\bibfield{author}{\bibinfo{person}{Yongye Su}, \bibinfo{person}{Yinqi Sun}, \bibinfo{person}{Minjia Zhang}, {and} \bibinfo{person}{Jianguo Wang}.} \bibinfo{year}{2024}\natexlab{}.
\newblock \showarticletitle{Vexless: A Serverless Vector Data Management System Using Cloud Functions}.
\newblock \bibinfo{journal}{\emph{Proceedings of the ACM on Management of Data}} \bibinfo{volume}{2}, \bibinfo{number}{3} (\bibinfo{year}{2024}), \bibinfo{pages}{1--26}.
\newblock


\bibitem[Sun et~al\mbox{.}(2023)]%
        {soar}
\bibfield{author}{\bibinfo{person}{Philip Sun}, \bibinfo{person}{David Simcha}, \bibinfo{person}{Dave Dopson}, \bibinfo{person}{Ruiqi Guo}, {and} \bibinfo{person}{Sanjiv Kumar}.} \bibinfo{year}{2023}\natexlab{}.
\newblock \showarticletitle{SOAR: improved indexing for approximate nearest neighbor search}.
\newblock \bibinfo{journal}{\emph{Advances in Neural Information Processing Systems}}  \bibinfo{volume}{36} (\bibinfo{year}{2023}), \bibinfo{pages}{3189--3204}.
\newblock


\bibitem[Vardanian(2023a)]%
        {simsimd}
\bibfield{author}{\bibinfo{person}{Ash Vardanian}.} \bibinfo{year}{2023}\natexlab{a}.
\newblock \bibinfo{booktitle}{\emph{SimSimd: Up to 200x Faster Dot Products \& Similarity Metrics}}.
\newblock
\urldef\tempurl%
\url{https://github.com/ashvardanian/SimSIMD}
\showURL{%
\tempurl}


\bibitem[Vardanian(2023b)]%
        {usearch}
\bibfield{author}{\bibinfo{person}{Ash Vardanian}.} \bibinfo{year}{2023}\natexlab{b}.
\newblock \bibinfo{booktitle}{\emph{{USearch by Unum Cloud}}}.
\newblock
\urldef\tempurl%
\url{https://doi.org/10.5281/zenodo.7949416}
\showDOI{\tempurl}


\bibitem[Wang et~al\mbox{.}(2021b)]%
        {milvus}
\bibfield{author}{\bibinfo{person}{Jianguo Wang}, \bibinfo{person}{Xiaomeng Yi}, \bibinfo{person}{Rentong Guo}, \bibinfo{person}{Hai Jin}, \bibinfo{person}{Peng Xu}, \bibinfo{person}{Shengjun Li}, \bibinfo{person}{Xiangyu Wang}, \bibinfo{person}{Xiangzhou Guo}, \bibinfo{person}{Chengming Li}, \bibinfo{person}{Xiaohai Xu}, {et~al\mbox{.}}} \bibinfo{year}{2021}\natexlab{b}.
\newblock \showarticletitle{Milvus: A purpose-built vector data management system}. In \bibinfo{booktitle}{\emph{Proceedings of the 2021 International Conference on Management of Data}}. \bibinfo{pages}{2614--2627}.
\newblock


\bibitem[Wang et~al\mbox{.}(2025)]%
        {graphmembounded}
\bibfield{author}{\bibinfo{person}{Mengzhao Wang}, \bibinfo{person}{Haotian Wu}, \bibinfo{person}{Xiangyu Ke}, \bibinfo{person}{Yunjun Gao}, \bibinfo{person}{Yifan Zhu}, {and} \bibinfo{person}{Wenchao Zhou}.} \bibinfo{year}{2025}\natexlab{}.
\newblock \showarticletitle{Accelerating Graph Indexing for ANNS on Modern CPUs}.
\newblock \bibinfo{journal}{\emph{arXiv preprint arXiv:2502.18113}} (\bibinfo{year}{2025}).
\newblock


\bibitem[Wang et~al\mbox{.}(2021a)]%
        {surveygraph}
\bibfield{author}{\bibinfo{person}{Mengzhao Wang}, \bibinfo{person}{Xiaoliang Xu}, \bibinfo{person}{Qiang Yue}, {and} \bibinfo{person}{Yuxiang Wang}.} \bibinfo{year}{2021}\natexlab{a}.
\newblock \showarticletitle{A comprehensive survey and experimental comparison of graph-based approximate nearest neighbor search}.
\newblock \bibinfo{journal}{\emph{arXiv preprint arXiv:2101.12631}} (\bibinfo{year}{2021}).
\newblock


\bibitem[Watts and Strogatz(1998)]%
        {smallworld}
\bibfield{author}{\bibinfo{person}{Duncan~J Watts} {and} \bibinfo{person}{Steven~H Strogatz}.} \bibinfo{year}{1998}\natexlab{}.
\newblock \showarticletitle{Collective dynamics of ‘small-world’networks}.
\newblock \bibinfo{journal}{\emph{nature}} \bibinfo{volume}{393}, \bibinfo{number}{6684} (\bibinfo{year}{1998}), \bibinfo{pages}{440--442}.
\newblock


\bibitem[Weaviate(2019)]%
        {weaviate}
\bibfield{author}{\bibinfo{person}{Weaviate}.} \bibinfo{year}{2019}\natexlab{}.
\newblock \bibinfo{booktitle}{\emph{{Weaviate}}}.
\newblock
\urldef\tempurl%
\url{https://github.com/weaviate/weaviate}
\showURL{%
\tempurl}


\bibitem[Wei et~al\mbox{.}(2025)]%
        {suco}
\bibfield{author}{\bibinfo{person}{Jiuqi Wei}, \bibinfo{person}{Xiaodong Lee}, \bibinfo{person}{Zhenyu Liao}, \bibinfo{person}{Themis Palpanas}, {and} \bibinfo{person}{Botao Peng}.} \bibinfo{year}{2025}\natexlab{}.
\newblock \showarticletitle{Subspace Collision: An Efficient and Accurate Framework for High-dimensional Approximate Nearest Neighbor Search}.
\newblock \bibinfo{journal}{\emph{Proceedings of the ACM on Management of Data}} \bibinfo{volume}{3}, \bibinfo{number}{1} (\bibinfo{year}{2025}), \bibinfo{pages}{1--29}.
\newblock


\bibitem[Xian et~al\mbox{.}(2024)]%
        {lucene}
\bibfield{author}{\bibinfo{person}{Jasper Xian}, \bibinfo{person}{Tommaso Teofili}, \bibinfo{person}{Ronak Pradeep}, {and} \bibinfo{person}{Jimmy Lin}.} \bibinfo{year}{2024}\natexlab{}.
\newblock \showarticletitle{Vector search with OpenAI embeddings: Lucene is all you need}. In \bibinfo{booktitle}{\emph{Proceedings of the 17th ACM International Conference on Web Search and Data Mining}}. \bibinfo{pages}{1090--1093}.
\newblock


\bibitem[Xu et~al\mbox{.}(2025)]%
        {tribase}
\bibfield{author}{\bibinfo{person}{Qian Xu}, \bibinfo{person}{Juan Yang}, \bibinfo{person}{Feng Zhang}, \bibinfo{person}{Junda Pan}, \bibinfo{person}{Kang Chen}, \bibinfo{person}{Youren Shen}, \bibinfo{person}{Amelie~Chi Zhou}, {and} \bibinfo{person}{Xiaoyong Du}.} \bibinfo{year}{2025}\natexlab{}.
\newblock \showarticletitle{Tribase: A Vector Data Query Engine for Reliable and Lossless Pruning Compression using Triangle Inequalities}.
\newblock \bibinfo{journal}{\emph{Proceedings of the ACM on Management of Data}} \bibinfo{volume}{3}, \bibinfo{number}{1} (\bibinfo{year}{2025}), \bibinfo{pages}{1--28}.
\newblock


\bibitem[Yang et~al\mbox{.}(2024)]%
        {bsa}
\bibfield{author}{\bibinfo{person}{Mingyu Yang}, \bibinfo{person}{Wentao Li}, \bibinfo{person}{Jiabao Jin}, \bibinfo{person}{Xiaoyao Zhong}, \bibinfo{person}{Xiangyu Wang}, \bibinfo{person}{Zhitao Shen}, \bibinfo{person}{Wei Jia}, {and} \bibinfo{person}{Wei Wang}.} \bibinfo{year}{2024}\natexlab{}.
\newblock \showarticletitle{Effective and General Distance Computation for Approximate Nearest Neighbor Search}.
\newblock \bibinfo{journal}{\emph{arXiv preprint arXiv:2404.16322}} (\bibinfo{year}{2024}).
\newblock


\bibitem[Zeng et~al\mbox{.}(2024)]%
        {candy}
\bibfield{author}{\bibinfo{person}{Xianzhi Zeng}, \bibinfo{person}{Zhuoyan Wu}, \bibinfo{person}{Xinjing Hu}, \bibinfo{person}{Xuanhua Shi}, \bibinfo{person}{Shixuan Sun}, {and} \bibinfo{person}{Shuhao Zhang}.} \bibinfo{year}{2024}\natexlab{}.
\newblock \showarticletitle{CANDY: A Benchmark for Continuous Approximate Nearest Neighbor Search with Dynamic Data Ingestion}.
\newblock \bibinfo{journal}{\emph{arXiv preprint arXiv:2406.19651}} (\bibinfo{year}{2024}).
\newblock


\end{thebibliography}

\appendix

\end{document}